\documentstyle[12pt]{article}

\def\be{\begin{equation}}
\def\ee{\end{equation}}
\def\bdi{\begin{displaymath}}
\def\edi{\end{displaymath}}
\def\br{\begin{eqnarray}}
\def\er{\end{eqnarray}}
\def\no{\nonumber}
\def\o{\over}

\def\u2{\mid u\mid^2}

\def\no{\nonumber}
\def\for{{\rm for}}

\def\RR{{\rm I\kern-.1567em R}}                              
 \def\CC{{\rm C\kern-4.7pt                                    
 \vrule height 7.7pt width 0.4pt depth -0.5pt \phantom {.}}} 
 \def\ZZ{{\sf Z\kern-4.5pt Z}}                                

\begin{document}

\begin{titlepage}
\vspace*{-2 cm}
\noindent

\vskip 3cm
\begin{center}
{\Large\bf Investigation of the Nicole model }
\vglue 1  true cm
C. Adam$^{a*}$,   J. S\'anchez-Guill\'en$^{a**}$,  
R.A. V\'azquez$^{a\dagger}$, and A. Wereszczy\'nski$^{b\dagger\dagger}$
\vspace{1 cm}

\small{ $^{a)}$Departamento de Fisica de Particulas, Universidad
       de Santiago}
       \\
       \small{ and Instituto Galego de Fisica de Altas Enerxias (IGFAE)}
       \\ \small{E-15782 Santiago de Compostela, Spain}
       \\ \small{ $^{b)}$Institute of Physics,  Jagiellonian
       University,}
       \\ \small{ Reymonta 4, 30-059 Krak\'{o}w, Poland}

\medskip
\end{center}

\normalsize
\vskip 0.2cm

\begin{abstract}
We study soliton solutions of the Nicole model - a non-linear
four-dimensional field theory
consisting of the $CP^1$ Lagrangian density
to the non-integer power $\frac{3}{2}$
- using an ansatz within
toroidal coordinates, which is indicated by the conformal symmetry of the
static equations of motion. We calculate the soliton energies numerically 
and find that they grow linearly with the topological charge (Hopf index).
Further we prove this behaviour to hold exactly for the ansatz.
On the other hand, for the full three-dimensional system without
symmetry reduction we prove a sub-linear upper bound, analogously to the case
of the Faddeev--Niemi model. It follows that symmetric solitons cannot
be true minimizers of the energy for sufficiently large Hopf index, again
in analogy to the Faddeev--Niemi model.

\end{abstract}

\vfill

{\footnotesize
$^*$adam@fpaxp1.usc.es

$^{**}$joaquin@fpaxp1.usc.es

$^\dagger$vazquez@fpaxp1.usc.es

$^{\dagger\dagger}$wereszczynski@th.if.uj.edu.pl }

\end{titlepage}

\section{Introduction }
In the last few years there has been rising interest in non-linear field 
theories which allow for the existence of ringlike, or, more generally,
knotlike solitons. On the one hand, this interest is due to the fact that
there may exist physical applications for such models, as is the case,
for instance, for the Faddeev--Niemi model, which finds some applications
both as a candidate for a low-energy effective theory for Yang-Mills theory
and in condensed matter physics. On the other hand, the rising interest
is related to the advance of more powerful computer facilities which allow
for reliable numerical calculations of those solitons in cases when an
analytical solution is not available (which happens quite often).
In addition, there is some intrinsical mathematical interest in 
theories with knot solitons. 

In the simplest case the field of the theory describes a map from one-point
compactified three-dimensional space $\RR_0^3$ to the two-sphere $S^2$.
$\RR_0^3$ is topologically equivalent to the three-sphere $S^3$, therefore
such maps are characterized by the third homotopy group of the target space
$S^2$, which is nontrivial,  $\pi_3 (S^2) =\ZZ$. As a consequence, fields
which describe maps $\RR_0^3 \to S^2$ fall into different homotopy classes,
and a soliton is a field configuration which minimizes a given energy
functional within a fixed homotopy class. The topological index 
characterizing the homotopy class is called Hopf index, the corresponding
map is called a Hopf map, and the minimizers are sometimes called  Hopf
solitons. For details on the Hopf map we refer to Appendix A.

The probably best-known theory which allows for Hopf solitons
is the Faddeev--Niemi model (\cite{Fad}, \cite{FN1}) with Lagrangian density
\be \label{FN-L}
{\cal L}_{\rm FN} = {\cal L}_2 - \lambda {\cal L}_4
\ee
where $\lambda $ is a dimensionful coupling constant, 
${\cal L}_2$ is 
\be \label{cp1}
{\cal L}_2 = 4 \frac{\partial_\mu u \, \partial^\mu \bar u}{(1+ u\bar u)^2} ,
\ee
and ${\cal L}_4$ is
\be
 {\cal L}_4 = 4 \frac{(\partial^\mu u \, \partial_\mu \bar u)^2 - (\partial^\mu
u \, \partial_\mu u)(\partial^\nu \bar u \, 
\partial_\nu \bar u)}{(1+u\bar u)^4} .
\ee
Further, $u$ is a complex field
which parametrizes the stereographic projection of the target $S^2$,
see Appendix A.
The Faddeev--Niemi model 
is the $S^2$ restriction of Skyrme theory and so
circumvents Derrick's theorem, because it consists
of two terms such that their corresponding energies behave oppositely under
a scale transformation. The existence of (static) soliton solutions for the 
lowest Hopf indices has been confirmed by numerical calculations 
(\cite{GH} -- \cite{HiSa}).

There are two more models which can be constructed
from the two Lagrangian
densities ${\cal L}_2$ and ${\cal L}_4$ separately by choosing appropriate
(non-integer) powers of these Lagrangians such that the corresponding energies
are scale invariant.
For ${\cal L}_4$ the appropriate choice is
\be
{\cal L}_{\rm AFZ} = -({\cal L}_4)^\frac{3}{4}
\ee
and for this model infinitely many analytic soliton solutions were found
by Aratyn, Ferreira and Zimerman (=AFZ) by using an ansatz with toroidal
coordinates (\cite{AFZ1}, \cite{AFZ2}). 
We shall, therefore, refer to this model as the AFZ model
in the sequel. The analysis of the AFZ model was carried further in 
(\cite{BF}), where, among other results, all the space-time and (geometric) 
target space symmetries of the AFZ model were determined, and, further,
the use of the ansatz with toroidal coordinates was related to the
conformal symmetry of the model (more precisely, of the static equations
of motion).  It turns out that
the AFZ model has infinitely many target space symmetries and, moreover,
realizes the notion of classical integrability in a rather strict sense,
because the static equations of motion (e.o.m.)
resulting from the ansatz with toroidal
coordinates may be solved by simple integration. Let us also mention here
that the model with Lagrangian ${\cal L}_4$ has the same (infinitely many)
target space symmetries and, in addition, a scale invariant action, leading
thereby to analytic time-dependent solutions \cite{Ferr2}.

The other model is
 \be \label{Ni-La}
{\cal L}_{\rm Ni}= ({\cal L}_2)^\frac{3}{2}.
\ee
This model has first been proposed by Nicole (\cite{Ni}), and it was shown
in the same paper that the simplest Hopf map with Hopf index 1 is a
soliton solution for this model. We, therefore, call this model the
Nicole (=Ni) model in this paper.
This model shares the conformal symmetry with the AFZ model, therefore,
again the ansatz with toroidal coordinates may be used to simplify the
static e.o.m. (to reduce them to an ordinary differential equation). 
However, the Nicole model only has the obvious symmetries - the conformal
base space symmetries (in the static case) and the modular target space
symmetries, see \cite{ASG2}. Consequently, the e.o.m. are no longer 
integrable, and the solutions are no longer available in closed,
analytic form (except for the simplest case with Hopf index 
one\footnote{Exact solitons with higher Hopf index have
been found in modified Nicole models \cite{nicole_modif}}). 
It is the main purpose of this paper to analyse these soliton solutions
with higher Hopf index by developing the analytical treatment as far as
possible and by performing numerical calculations where it is
necessary.

The AFZ and Nicole models currently do not have direct physical applications,
but they can serve as useful test labors for the understanding of
generic features of models with knot solitons.   
This is all the more true as analytic results - already for solitons, but even
more so for time-dependent problems like soliton scattering - are notoriously
difficult to obtain for realistic models like the Faddeev--Niemi model.
Both the AFZ and the Nicole model are much easier to treat, and for time 
dependent problems the Nicole model even has one advantage over the AFZ
model. The infinitely many symmetries of the former - which provide its
integrability on the one hand - imply that the moduli
space for multi-soliton configurations is infinite dimensional. This 
most likely makes an adiabatic treatment of soliton scattering problematic.
On the other hand, that problem is absent for the Nicole model which
may, therefore, probably serve as a natural test labor for an
adiabatic study of the scattering of knot-like solitons. A further issue which
may be useful for the study of time dependent problems is provided by
the observation that all field configurations within the toroidal ansatz
belong to an integrable subsector of the Nicole model (i.e., to a sector
with infinitely many conservad currents), see Section 3.

The paper is organized as follows. In Section 2 we provide the static
e.o.m. of the Nicole model as well as the non-linear
ODE resulting from the ansatz
with toroidal coordinates. Then we analyse the latter  equation in detail,
establishing some properties analytically. In a next step we calculate the 
energies of the corresponding solitons numerically with high precision. 
We provide all energies for solitons 
with sufficiently small Hopf index. Further, we calculate some energies
for larger Hopf index in order to establish some generic behaviour.
We prove that
this generic behaviour is in accordance with exact energy estimates for
the ansatz in toroidal coordinates.

In Section 3 we show that for all the solutions of Section 2 there exist
infinitely many conserved currents, that is to say, all these field
configurations belong to an integrable subsector of the Nicole model.

In Section 4.1 we prove an upper bound for the energies of field configurations
with given Hopf index, of the type $ E_Q \le C_2 Q^\frac{3}{4}$, and
provide an explicit value for the constant $C_2$. 
In Section 4.2 we derive an analogous bound for the AFZ and Faddeev--Niemi 
model, which is easily achieved with the help of the results of Section 4.1.
The discussion of
the possibility to find a corresponding lower bound $E_Q \ge C_1 
Q^\frac{3}{4}$ for the Nicole model, 
analogously to the Vakulenko-Kapitansky bound for the
Faddeev--Niemi model, is presented in Appendix C. 
We show where a Vakulenko--Kapitansky type
proof needs to be refined for the Nicole (and also for the AFZ) model and 
leave this lower bound as a conjecture for the moment.

Section 5 contains our conclusions, where  
we comment on the relevance of our results for issues like
stability, and on the relation to the corresponding
results for the other two models (Faddeev--Niemi and AFZ). 
In Appendix A we collect useful
results and facts about Hopf maps which we use throughout the paper.
Appendix B contains some proofs which we need in the main sections.
In Appendix C we discuss the lower energy bound.

\section{Solitons in the Nicole model}

\subsection{Equation of motion}

The static e.o.m. for the Nicole model (\ref{Ni-La}) is
\be \label{Ni-eom}
\frac{1}{2}(u_{jk}\bar u_j u_k + \bar u_{jk}u_j u_k) +u_j \bar u_j u_{kk}
-\frac{u_j \bar u_j}{1+u\bar u}(u_k \bar u_k u +3 u_k u_k \bar u)=0
\ee
where $u_k \equiv \partial_\mu u$ etc., and the Einstein summation 
convention is understood. We now introduce toroidal coordinates
$(\eta ,\xi ,\varphi)$ via
\br
x &=&  q^{-1} \sinh \eta \cos \varphi \;\;, \;\;
y =  q^{-1} \sinh \eta \sin \varphi   \nonumber \\
z &=&  q^{-1} \sin \xi \quad ;  \qquad  q = \cosh \eta - \cos \xi .
\label{tordefs}
\er
Further, we choose for $u$ the ansatz
\be \label{u-mn}
u = f(\eta)\, e^{ im\varphi + in\xi} \quad , \quad m,n \in \ZZ
\ee
which is compatible with the e.o.m. for the Nicole model as well as for the
AFZ model. The reason for this ansatz may, in fact, be understood from
the symmetries of the model. More precisely, 
the presence of the conformal symmetry on base space implies that the
ansatz (\ref{u-mn}) is an ``educated guess'' for a solution to Eq.
(\ref{Ni-eom}) in the sense of the Lie theory of symmetry. That is
to say, if we choose a rotation about the $z$ axis and a certain
combination of proper conformal transformation along the $z$ axis and 
translation along the $z$ axis as a maximal set of two commuting base 
space transformations, then the corresponding infinitesimal 
symmetry generators
(vectors ${\bf v}^i$) are precisely given by the tangent vectors along
$\varphi$ and $\xi$, ${\bf v}^1 =\partial_\varphi$, and ${\bf v}^2 =
\partial_\xi $. The ansatz (\ref{u-mn}) is invariant under a combination
of these base space transformations and  phase transformations of the 
target space variable $u$, i.e., under the action of the vector
fields $\tilde {\bf v}^1 = \partial_\varphi -im u \partial_u$ and
$\tilde {\bf v}^2 = \partial_\xi -in u \partial_u$, which provides
precisely the educated guess according to Lie. A concise discussion of
these points can be found in Ref. \cite{BF}, 
where the symmetries of the AFZ model
are discussed in great detail.

We need  the gradient in toroidal coordinates
 \be \label{grad-3}
\nabla = (\nabla \eta)\partial_\eta 
+(\nabla \xi )\partial_\xi +(\nabla \varphi)\partial_\varphi
= q(\hat e_\eta \partial_\eta + \hat e_\xi \partial_\xi +
\frac{1}{\sinh \eta} \hat
e_\varphi \partial_\varphi  )
\ee
where $(\hat e_\eta  ,\hat e_\xi ,\hat e_\varphi )$ form an orthonormal
frame in $\RR^3$.  Further we need the relations
\be
\nabla \cdot \hat e_\eta = -\sinh \eta + \frac{1-\cosh \eta \cos 
\xi}{\sinh \eta} \, ,\quad \nabla \cdot \hat e_\xi = -2 \sin \xi \, ,\quad
\nabla \cdot \hat e_\varphi =0 .
\ee
Inserting now the ansatz (\ref{u-mn}) into the static e.o.m. 
(\ref{Ni-eom}) we find, after a straight-forward calculation,  
that the ansatz is indeed compatible with the static e.o.m. and that $f$ 
has to obey the non-linear ODE
\begin{displaymath}
2 f'' f'^2 + f'' f^2 \left( n^2 + \frac{m^2}{\sinh^2 \eta } \right) 
+ \left( f'^3 + n^2 f' f^2 \right) 
\frac{\cosh \eta}{\sinh \eta } 
 - f^3 \left( n^2 + \frac{m^2}{\sinh^2 \eta } \right)^2 - 
\end{displaymath}
\be
- f \frac{f'^2 + f^2 \left( n^2 + \frac{m^2}{\sinh^2 \eta } \right) }{1+f^2}
\left( 4f'^2 - 2f^2 \left( n^2 + \frac{m^2}{\sinh^2 \eta } \right) 
\right) =0
\ee
where $f' \equiv \partial_\eta f$ etc. It may be checked without difficulty 
that the simplest Hopf map 
\be \label{s-ho}
u=\sinh \eta \, e^{i\xi + i \varphi } \quad \Rightarrow \quad
f =\sinh \eta \, , \quad
m=n=1
\ee
indeed solves the above equation.
 
For technical reasons it is preferable to introduce the new variable
\be
t\equiv \sinh \eta
\ee
in terms of which the above e.o.m. can be expressed as a pure polynomial 
in the independent
and dependent variables $t,f,f_t ,f_{tt}$,
\bdi
F(t,f,f_t,f_{tt}) \equiv
\edi
\bdi
 t^2 (1+t^2)(1+f^2) \left( 2t^2 (1+t^2) f_t^2 + (n^2 t^2 +m^2) 
f^2 \right) f_{tt} -4 t^4 (1+t^2 )^2 f f_t^4 + 
\edi
\bdi
+ t^3 (1+3t^2)(1+t^2) (1+f^2) f_t^3 - 2t^2 (1+t^2) (n^2 t^2 +m^2) f^3 f_t^2 +
\edi
\be \label{eom-t}
+ t^3 \left( m^2 + n^2 (1+2t^2)\right) (1+f^2) f^2 f_t - (n^2 t^2 +m^2 )^2
f^3 (1-f^2) =0
\ee   
where $f_t \equiv \partial_t f$ etc. Further, we need the energy functional
$E[f]$ which results for the ansatz (\ref{u-mn}) after the integration
w.r.t. the variables $\xi ,\varphi $. With
\be
dV \equiv d^3 r = q^{-3} \sinh \eta \, d \eta \, d\xi \, d\varphi
\ee
we find
\br \label{e-func}
E[f] &=& 32 \pi^2 \int_0^\infty d\eta \sinh \eta \frac{ \left( f_\eta^2 + f^2
\left( n^2 + \frac{m^2}{\sinh^2 \eta} \right) \right)^\frac{3}{2}}{
(1+f^2 )^3} \nonumber \\
&=& 32\pi^2 \int_0^\infty dt \, t (1+t^2) \frac{ \left( f_t^2 + \left( n^2 +
\frac{m^2}{t^2} \right) \frac{f^2}{1+t^2} \right)^\frac{3}{2}}{
(1+f^2)^3}
\er  
For $n=m=1$ for the solution $f(t) =t$ the energy may be calculated 
analytically,
\be
E=32\pi^2 2^\frac{3}{2} \int_0^\infty \frac{dt t}{(1+t^2)^2} =32 \pi^2 
\sqrt{2} .
\ee
If we calculate the energy for the field configuration $f(t)= ct$
instead, we find 
\be
E=  32 \pi^2  \sqrt{2} \frac{1}{2} (c+\frac{1}{c})
\ee
which certainly has a minimum at $c=1$ but also shows clearly that the
one-soliton sector is indeed separated from the trivial sector (with Hopf 
index zero) by an infinite energy barrier.

Next we have to fix the boundary conditions which $f$ has to obey. The
$u$ of Eq. (\ref{u-mn}) has to take values in the whole target space
$\CC_0$ in order to be a Hopf map. This implies that $f$ has to take values 
in the whole positive real numbers, including zero, i.e., in the  whole
$\RR_{+0}$. It follows that the possible boundary conditions on $f$ are
restricted to one of the two following options. Either
$f(0)=0$, $f(\infty) =\infty $ or $f(0)=\infty $, $f(\infty) =0$. The
reason for this is that the values $f=0$ and $f=\infty$ correspond to
the north and south pole of the target space $S^2$, respectively.
Consequently, their pre-images in the base space $\RR_0^3$ must be
one-dimensional objects (closed lines). But the only two values 
$\eta =\eta_0 = {\rm const}$ (or $t=t_0 ={\rm const}$) 
which correspond to lines rather than surfaces (tori) are the values
$\eta =0$ ($t=0$) and $\eta =\infty$ ($t=\infty$). Therefore we assume
\be 
f(\eta =0) =f(t=0) =0 \, ,\quad f(\eta =\infty )=
 f(t=\infty )=\infty
\ee
which is general, because the other option is related to this one by a 
symmetry transformation.   

\subsection{Asymptotic behaviour and numerical evaluation}

In a next step we want to determine the asymptotic behaviour of the function 
$f(t)$ for small and large values of $t$ from the differential equation
(\ref{eom-t}).  For small $t$ we assume that $f(t) \sim t^\alpha +
o(t^{ \beta})$ where $\beta > \alpha>0$. From Eq. (\ref{eom-t}) we get
\be
F(t,f,f_t ,f_{tt}) \sim (2 \alpha^4 -\alpha^3 + \alpha^2 m^2 - \alpha m^2 -
m^4) t^{3 \alpha}  + o(t^{3\alpha +2}) +o(t^{3 \beta}) \equiv 0
\ee
The condition that the leading (i.e., smallest) 
order $t^{3\alpha}$ is absent therefore leads to
\be
2 \alpha^4 -\alpha^3 + \alpha^2 m^2 - \alpha m^2 - m^4 =0
\ee
which has four solutions for $\alpha$ of which only one is acceptable
(real and positive), namely
\be
\alpha_m \equiv \frac{1}{4} \left( 1 + \sqrt{8 m^2 +1} \right)
\ee
In the same fashion we may determine the subleading (higher order)
contributions to $f$ for small $t$ in an iterative manner. We find
\be
f = t^{\alpha_m} P^{(0)}(t ,t^{\alpha_m}) \qquad {\rm for}\qquad
t << 1
\ee
where $P^{(0)}(t ,t^{\alpha_m})$ is a polynomial of its arguments, which is
determined up to a multiplicative constant,
\be
P^{(0)}(t ,t^{\alpha_m}) =\sum_{k,l=0}^\infty c^{(0)}_{kl}t^k t^{l\alpha_m} .
\ee
Here it is possible to determine the higher coefficients $c^{(0)}_{kl}$ 
in terms of $m$, $n$ and $c^{(0)}_{00}$. Unfortunately, it is not possible
to determine  $c^{(0)}_{00}$, i.e., an overall constant remains undetermined
by this asymptotic analysis.
  
We repeat the same asymptotic analysis for large values of $t$. We assume
that $f(t) \sim t^\alpha + o(t^{ \beta})$ where now $\alpha >\beta >0$
and find
\be
F(t,f,f_t ,f_{tt}) \sim (2 \alpha^4 -\alpha^3 + \alpha^2 n^2 - \alpha n^2 -
n^4) t^{5 \alpha +4}  + o(t^{5\alpha +2}) +o(t^{5 \beta +4}) \equiv 0
\ee
and therefore the same condition for $\alpha$ 
like for small t with the only replacement
$m\to n$. We may again determine $f$ asymptotically up to an overall
constant,
\be
f = t^{\alpha_n} P^{(\infty )}(t^{-1} ,t^{-\alpha_n}) \qquad {\rm for}\qquad
t >> 1
\ee
where 
\be
P^{(\infty )}(t^{-1} ,t^{-\alpha_n}) 
=\sum_{k,l=0}^\infty c^{(\infty )}_{kl}t^{-k} t^{-l\alpha_n}
\ee
and again the higher coefficients $c^{( \infty )}_{kl}$ are determined
in terms of $m$, $n$ and $c^{(0)}_{00}$. Again, $c^{(\infty )}_{00}$
remains undetermined.

Next we turn to the numerical evaluation of the soliton energies for general
$m,n$. For this purpose it is preferable to minimize the energy functional 
(\ref{e-func}). The problem is that standard evolution procedures for the
differential equation (\ref{eom-t}) cannot be used because of the singular
nature of this equation at $t=0$. We proceed as follows with the 
minimisation. We first factorize the leading behaviour, e.g., we choose
\be
f(t) \equiv t^{\alpha_m} (1+t^2)^{\frac{\alpha_n - \alpha_m}{2}} g(t)
\ee
where $g$ obeys
\be
g>0 \quad \forall \quad t \, , \quad g(0)= g_0 \, , \quad g(\infty )
= g_\infty 
\ee
Then we make a finite parameter ansatz for $g$ of the form
\be
g (t)= \frac{g_0 + g_\infty t^2}{1+t^2} \, \frac{1 + \sum_{i=1}^k
a_i t^{2i} + t^{k+2}}{1+ \sum_{j=1}^k b_j t^{2j} + t^{k+2}}
\ee
where $g_0 $, $g_\infty$, $a_i$ and $b_j$ are the parameters with respect
to which we minimize the energy functional (\ref{e-func}). For each value of
$m,n$ we increase the number of parameters (i.e., the integer number $k$),
until we reach stability (i.e., until the energies do not change further with 
the increase of the number of parameters). We do not choose a full
numerical minimization of the energy functional for the following reason.
It turns out that the leading behaviour (i.e., $g(t) = {\rm const}$),
already provides rather good results for the corresponding energies.
This implies that the energy functional, when viewed as a functional
of $g$, is rather shallow. As a consequence, a full numerical minimisation
which implements the derivatives numerically has the tendency to 
produce false minima unless the numerical minimisation grid is chosen
very finely (and consumes a lot of computer time). On the other hand,
within our finite parameter
ansatz we are able to perform the derivatives analytically,
and therefore this problem is absent. 

We show our numerical results in Table 1 for all $(m,n)= (1,1) \ldots (5,5)$ 
as well as for some higher values. Further we plot the energies versus the
Hopf index $Q=mn$ in Fig. 1. From our results it is obvious that the 
energies grow linearly in $Q$ to a very good accuracy. 
More precisely, the energies for $m=n$ lie on a straight line in Fig. 1 almost
exactly. The energies for $m\not= n$  lie slightly above, whereas no energy 
lies below this line. It is in fact possible to prove the linear growth 
of energies with the Hopf charge for solutions of the ansatz  (\ref{u-mn}).
This we show for the simplest case $m=n$ in the next subsection.

On the other hand, for the full model 
the sublinear upper bound $E\le C_2 Q^\frac{3}{4}$ holds, which we prove in 
Section 4. It follows that the solutions we found for the symmetric ansatz
(\ref{u-mn}) {\em cannot} be true solitons, i.e., global minima of topological
sectors with fixed $Q$ for sufficiently large values of $Q$. This is akin to 
the situation in the Faddeev--Niemi model and differs from the situation in the
AFZ model (see our discussion in Section 5). 

\begin{table}
\begin{tabular}{||r|r|c||r|r|c||}
\hline
$n$ & $m$ & $ E^*$ &$n$ & $m$ & $ E^*$ \\
\hline
 1& 1&  1.41362193  & 4& 1&  6.52450266\\ 
 1& 2&  2.62615091 &  4& 2&  9.03384901\\ 
 1& 3&  4.13338723 &  4& 3&  12.7436624\\ 
 1& 4&  5.37349409 &  4& 4&  16.789796\\ 
 1& 5&  7.54150789 &  4& 5&  20.886948\\ 
 2& 1&  2.64753548 &  5& 1&  9.48680075\\ 
 2& 2&  4.52249973 &  5& 2&  11.8901415\\ 
 2& 3&  6.62925943 &  5& 3&  15.9786399\\
 2& 4&  8.85370929 &  5& 4&  20.8937915\\
 2& 5&  10.9513184 &  5& 5&  25.975942\\  
 3& 1&  4.28232266 &  6& 6 & 37.2017869 \\ 
 3& 2&  6.64941801 &  8& 8&  65.771712  \\  
 3& 3&  9.64059435 &  9& 9&  83.11675 \\ 
 3& 4&  12.7307136 & 10& 10&  102.501125\\
 3& 5&  15.8672836 &  20& 20&  408.581244\\ 
 & &   &  5& 10& 51.8990747 \\
 & &   &  6& 15&  96.0585933\\
 & &   &  14& 7&  103.670713\\
\hline
\end{tabular}
\caption{The rescaled energies $E^* = (32 \pi^2)^{-1} E $
of solutions for selected values of $m$ and $n$.}
\end{table}

\subsection{Upper and lower bounds for the torus ansatz}

In this subsection 
we prove the bounds $ c_1 Q \le E [f_{m,n}]  $ and, for the special 
case $m=n$, $ E [f_{m,m}] \le c_2 m^2 $ (with explicit 
values for the constants $c_1$ and $c_2$) in order to
further support our numerical results. Here $E [f]$
is the energy functional (\ref{e-func}) 
and $f_{m,n}$ is the corresponding
solution of the equation of motion.  We first prove the lower bound.
We use the inequality 
\be
(a+b)^\frac{3}{2} \ge \frac{3\sqrt{3}}{2} a^\frac{1}{2} b \qquad
{\rm for } \qquad a,b\ge 0
\ee
which we prove in Appendix B (here $\frac{3\sqrt{3}}{2}$ is an optimal
value). We 
choose $a=f_t^2$ and $b=(n^2 + m^2/t^2 )f^2/(1+t^2)$ and 
introduce
\be 
Q=mn \, , \quad \mu =\frac{m}{n}
\ee
to get
\br
E [f] &=& 32 \pi^2 \int_0^\infty dt t (1+t^2) 
\frac{\left( f_t^2 + Q\left( \frac{1}{\mu} + \frac{\mu}{t^2} \right) 
\frac{f^2}{1+ t^2} \right)^\frac{3}{2}}{(1+f^2)^3} \nonumber \\
&\ge & 32 \pi^2 Q \frac{3\sqrt{3}}{2} \int_0^\infty dt \left(
\frac{t}{\mu} +\frac{\mu}{t} \right)
|f_t |
\frac{f^2}{(1+f^2)^3} \nonumber \\
&\ge & 32 \pi^2 Q \, 3\sqrt{3} \int_0^\infty dt f_t
\frac{f^2}{(1+f^2)^3} \equiv \tilde E[f]
\er
where we used $(t/\mu +\mu /t) \ge 2$. 
Up to now we have in fact used inequalities 
for the energy densities, i.e., these inequalities hold for given, fixed
functions $f$. However, if we evaluate the last expression for its minimizer
$f_{\rm min}$ then the inequality holds for arbitrary $f$ and, consequently,
for the solutions $f_{m,n}$. The minimization of the last expression is 
simplified by the observation that it is a total derivative. Therefore it
is minimized for all $f$ which obey the required boundary conditions 
 $f(0)=0$ and $f(\infty )=\infty$.
Explicitly we get
\br
\tilde E[f] &=& 32 \pi^2 Q \, 3\sqrt{3} \left[ -\frac{f}{4(1+f^2)^2} + \frac{f
}{8(1+f^2)} +\frac{1}{8} \arctan f \right]_{f=0}^{f=\infty} 
\nonumber \\
&=& 32 \pi^2 Q \, 3\sqrt{3} \frac{\pi}{16} =1.02 \times 32 \pi^2 Q.
\er
Comparing with Table 1, we see that for larger values of $m,n$ our lower 
bound gets very close to the numerical values, especially for $m=n$ 
(e.g., better than one per cent
for $m=n=10$ or $m=n=20$). This demonstrates the very good accuracy of our
numerical results.

For the derivation of an upper bound we restrict to the simpler case
$m=n$. We simply insert a trial function
$f_{m,m}^{\rm tr}$ into the energy functional $E [f]$. Concretely
we choose $f_{m,m}^{\rm tr} =t^m$ and get
\be
E [t^m] = 32 \pi^2 \int_0^\infty dt t (1+t^2) \frac{m^3 t^{3m-3}
2^\frac{3}{2}}{(1+t^{2m})^3} 
\ee
and, with the substitution $u=t^m$,
\br \label{trial-fm}
E [t^m] &=& 32 \pi^2 m^2 2^\frac{3}{2} \int_0^\infty du \left( u^\frac{1}{m}
+ u^{-\frac{1}{m}} \right) \frac{u^2}{(1+u^2)^3} \nonumber \\
&=& 32 \pi^2 m^2 2^\frac{3}{2} \frac{1}{8}\left( 1-\frac{1}{m^2} \right) 
\frac{\pi}{\cos \frac{\pi}{2m}} .
\er
The function $(1-1/m^2)/(\cos \frac{\pi}{2m})$ is a monotonously decreasing
function in $m$ and may, therefore, be estimated by its value at $m=2$
for the range $m\ge 2$.
With
\be
\left( 1-\frac{1}{4} \right) 
\frac{1}{\cos \frac{\pi}{4}} = \frac{3}{4}\sqrt{2}
\ee
we therefore get
\be \label{trial-m2}
E [t^m] \le 32 \pi^2 m^2 \frac{3\pi}{8} = 1.18 \times 32 \pi^2 m^2
\quad {\rm for} \quad m\ge 2
\ee
which, for large $m$, is about 15 percent above the numerical value.
By using Eq. (\ref{trial-fm}) instead of the order $m^2$ estimate 
(\ref{trial-m2}), we get
slightly better results which, for large $m$, are less than 10 percent
above the numerical values. 

In short, we established the bounds $c_1 m^2 \le E [f_{m,m}] \le c_2 m^2$
with $c_1 =1.02 \times 32 \pi^2$ and $c_2 = 1.18 \times 32 \pi^2$, as
announced.

\section{Infinitely many conserved currents}

In this section we want to demonstrate that the ansatz (\ref{u-mn}) belongs
to a subsector of the Nicole model with infinitely many conserved currents
for arbitrary profile function $f(\eta)$. The integrability condition
defining this subsector exists for a wide class of models, 
and discussing it for the whole class does not complicate matters
(for a more detailed discussion, and especially for the geometric meaning
of the integrability condition, we refer to \cite{ASGW2}; the concept of
higher dimensional integrability, which is at the basis of these discussions,
was introduced in \cite{AFSG}).
Therefore, we start with
the general class of Lagrangian densities
\be \label{g-lan}
{\cal L} (u ,\bar u ,u_\mu ,\bar u_\mu ) = F(a,b,c)   
\ee
where
\be
a=u\bar u \, ,\quad b=u_\mu \bar u^\mu \, ,\quad c= (u_\mu \bar u^\mu )^2
- u_\mu^2 \bar u_\nu^2 
\ee
and $F$ is an arbitrary real function of its arguments. Obviously,
the Nicole model belongs to this class, ${\cal L}_{\rm Ni} =
[4(1+a)^{-1} b]^\frac{3}{2}$. 

Further, we define the currents 
\be \label{k-mu}
K^\mu = h(a) \bar \Pi^\mu
\ee
where $h$ is an arbitrary given real function of its argument, and
$\Pi^\mu$ and $\bar \Pi^\mu$ are the conjugate four-momenta of
$u$ and $\bar u$, i.e.,
\be
\Pi_\mu \equiv {\cal L}_{u^\mu} = \bar u^\mu F_b + 2 (u^\lambda \bar
u_\lambda \bar u_\mu - \bar u_\lambda^2 u_\mu )F_c.
\ee
Finally, we define the infinitely many Noether currents
\be \label{noet-cu}
J^G_\mu = i ( G_u K_\mu  - G_{\bar u} \bar K_\mu ) 
\ee
where $G$ is an arbitrary real function of $u$ and $\bar u$, and
$G_u \equiv \partial_u G$. 

Now we want to study the conditions which one has to impose in order to make
the divergence  $\partial^\mu J^G_\mu$ vanish. A simple calculation reveals
\br \label{div-jg}
\partial^\mu J^G_\mu & = & i h \left( [( \frac{h'}{h} \bar u G_u + G_{uu} ) 
u_\mu^2 
 - ( \frac{h'}{h} u G_{\bar u} + G_{\bar u\bar u}) \bar u_\mu^2 ] F_b  \right.
\nonumber \\
&& \left. + \, (uG_u - \bar u G_{\bar u}) 
[ \frac{h'}{h} (bF_b + 2 cF_c ) +F_a ] \right) 
\er
where the prime denotes the derivative with respect to $a$.

The second term at the r.h.s. of Eq. (\ref{div-jg}) certainly
vanishes if
\be \label{cond1}
uG_u - \bar u G_{\bar u} =0
\ee
with the general solution
\be \label{cond1a}
G(u ,\bar u) = \tilde G (u\bar u) \equiv \tilde G(a) .
\ee
Without taking condition (\ref{cond1a}) into account, the 
first term on the r.h.s. of Eq. (\ref{div-jg}) vanishes for the integrability
condition 
\be \label{eik-eq}
u_\mu^2 =0
\ee
i.e., the complex eikonal equation (it also vanishes for
some more complicated integrability conditions which we do not discuss
here, see Refs. \cite{ASG3}, \cite{Wer3}). However, by
using condition (\ref{cond1a}) we may re-express the first term like
\be
(h' G' + h G'' )F_b [\bar u^2 u_\mu^2 - u^2 \bar u_\mu^2] 
\ee
and, therefore, we find, instead of the complex eikonal equation, the
weaker integrability condition
\be \label{int-c}
 \bar u^2 u_\mu^2 - u^2 \bar u_\mu^2 =0.
\ee
The meaning of this condition becomes especially transparent when we
re-express $u$ in terms of its modulus and phase like
\be
u=\exp (\Sigma + i \Lambda ).
\ee
Then the complex eikonal equation is equivalent to the two real equations
\be \label{re-c1}
\Sigma_\mu^2 = \Lambda_\mu^2
\ee
and
\be \label{re-c2}
\Sigma^\mu \Lambda_\mu =0  
\ee
whereas the weaker condition (\ref{int-c}) becomes Eq. (\ref{re-c2}) alone
or, for time-independent $u$, 
\be \label{re-c2a}
(\nabla \Sigma) \cdot (\nabla \Lambda )=0.
\ee

{\em Remark:} The condition $G (u ,\bar u) = \tilde G(u\bar u)$, Eq.
(\ref{cond1a}), restricts 
the space of allowed $G$ to a subspace which is still infinite-dimensional, 
therefore
the integrability condition (\ref{int-c}) really defines a subsector
with infinitely many conserved currents.

{\em Remark:} 
The ansatz (\ref{u-mn}) obeys the integrability condition (\ref{re-c2a})
for arbitrary profile function
$f(\eta)$, as is obvious from the orthonormality of the basis vectors
$(\hat e_\eta ,\hat e_\xi ,\hat e_\varphi )$. Therefore, all our solutions
of Section 2 really belong to the subsector of the Nicole
model defined by condition
(\ref{int-c}) with infinitely many conserved currents. This is in contrast 
to the complex eikonal equation (\ref{eik-eq}), which leads to a non-linear
ODE for the profile function which is not compatible with the Nicole model
field equation except for the simplest Hopf map $m=n=1$, see \cite{Ada1}.

{\em Remark:} Both condition (\ref{cond1a}) and the integrability condition
(\ref{int-c}) do not restrict the Lagrangian density, therefore the
corresponding 
integrable subsectors with infinitely many conservation laws exist for
all Lagrangians of the type (\ref{g-lan}). So they may be of interest
for other models like, e.g., the Faddeev--Niemi model.

\section{Energy estimates for the full model }

In this section we study  upper bounds on the energies of
solitons with a given Hopf index, of the type $
E_Q \le C_2 Q^\frac{3}{4}$. In Subsection 4.1 we prove the upper bound
$E_Q \le C_2 Q^\frac{3}{4}$ by following the method used in \cite{LiYa}
for the Faddeev--Niemi model. Further, we provide an explicit value
for the constant $C_2$. In Subsection 4.2, we briefly derive the
analogous upper bounds for the AFZ and the Faddeev--Niemi models,
again with explicit values for the constants $C_2$. They easily follow 
from our results from Subsection 4.1. 
The issue of a lower bound $C_1 Q^\frac{3}{4} \le E_Q$ is discussed in 
Appendix C. There we show
where a naive Vakulenko--Kapitansky type proof does not work and needs
some refinement in our case, leaving open for the moment
the problem of rigorously establishing such a lower bound. 
  
\subsection{The estimate $E_Q \le C_2 Q^{3/4}$ }

In this section we use the unit vector $\vec n$ to denote Hopf maps, because 
this has some technical advantages (components of the vector $\vec n$ can
be easily estimated, whereas the complex function $u$ may take arbitrary
values). In our proof we will closely follow the method used in 
\cite{LiYa} for the proof of an analogous result for the Faddeev--Niemi model.
However, we will use explicit expressions for certain functions instead
of using just their existence, which enables us to provide an explicit value
for the constant $C_2$ in our estimate.

For a Hopf index which is a square, $Q=l^2$, a Hopf map is constructed in
\cite{LiYa} which is a composition of a Hopf map 
$\vec n(\vec r)$ with Hopf index 1 and
a map $\vec N (\vec n)\, : \, S^2 \to S^2$ 
with winding number $w=l$. The Hopf map $\vec n (\vec r)$ 
is constructed such that it varies smoothly from $\vec n = (0,0,1)$ at 
$\vec r =0$ to $\vec n =(0 ,0,-1)$ at $|\vec r|=R$ and remains at this
value for $r > R$. This implies, by Eq. (\ref{q-s3-r}) of Appendix A,
that it has indeed Hopf index 1.
Further, the gradient $|\nabla^{\RR^3} \vec n|$ 
may be estimated like $|\nabla^{\RR^3} \vec n| \le c/R$ for $r\le R$ and by
$|\nabla^{\RR^3} \vec n| =0$ for $r > R$ (here and below
$c$ denotes an unspecified constant). The map $\vec N (\vec n):\, 
S^2 \to S^2$ is constructed such that $\vec N$ varies smoothly over $l$
nonintersecting geodesic discs covering the base $S^2$, where 
$\vec N =(0,0,1) $ in the center
of each disc and $\vec N =(0,0,-1)$ at the boundary of each disc and between
the discs. Each disc contributes a winding number 1 for the map 
$\vec N(\vec n)$, therefore the total winding number is $l$. Further, the
geodesic radius of each disc can be chosen (not larger than) 
$R_g = cl^{-\frac{1}{2}}$ in order to be able to put $l$ nonintersecting
discs on one unit $S^2$. The gradient may be again estimated by the
inverse (geodesic) radius, $|\nabla^{S^2}\vec N(\vec n) |
\le cl^{\frac{1}{2}}$,
and with the help of the chain rule one finds for the gradient of $\vec N$
that $|\nabla^{\RR_0^3}\vec N(\vec n(\vec r)) | \le c R^{-1}l^{\frac{1}{2}}$ 
and for the energy density ${\cal E}_{\rm Ni}(\vec r) 
\sim |\nabla^{\RR_0^3} \vec N|^3
\le c R^{-3}l^{\frac{3}{2}}$. The energy $E_{\rm Ni}$, 
which is the integral over all
space of ${\cal E}_{\rm Ni} (\vec r)$, 
may in the case at hand be calculated by integrating
over the ball $B^R$ of radius $R$ with the result that
$E_{\rm Ni} \le cl^\frac{3}{2} \equiv cQ^{\frac{3}{4}}$ which is the announced 
estimate for $Q=l^2$. 

Before rederiving this result in more detail and with explicit
choices for the maps described above in rather general terms, it may be
useful to have a geometric picture of the above compostion of maps.
In fact, the map $\vec N (\vec n (\vec r))$ describes $l$ nonintersecting,
 full tori
confined to the ball $B^R$, where, in addition, each full torus is linked
with all other tori. In the center of each torus there is a closed line along 
which $\vec N$ takes the value $\vec N=(0,0,1)$. From the center to the
surface of each full torus, $\vec N$ varies smoothly from 
$\vec N =(0,0,1)$ to $\vec N = (0,0,-1)$. 
At the surface of each full 
torus (which has the topology of the two-torus $T^2$) as well as in the space
between the tori and outside the ball $B^R$, $\vec N$ remains at the value
$\vec N=(0,0,-1)$. 

Now for the detailed derivation, where we still assume $Q=l^2$
for the moment. For the map
$\vec n (\vec r) =(\sin \vartheta \cos \phi ,\sin \vartheta \sin \phi ,
\cos \vartheta )$ with Hopf index 1
we assume that it has cylindrical symmetry, as in Eq. 
(\ref{n-rs1}) of Appendix A. For the profile function $\rho (r)$ we want
to find a choice which varies as smoothly as possible from $0$ to $\pi$ as
$r$ varies from $0$ to $R$, in order to be able to estimate it with a
number as small as possible. The smoothest choice $\rho = \pi r/R $ for 
$ r\le R$ and $\rho =\pi$ for $r>R$ does not have a continuous 
first derivative at $r=R$,
which we need for the energy density, but there is a simple generalization 
which does  have one. We choose
\be \label{rho-e}
\rho_\epsilon (r) =\pi F_\epsilon (x) \, ,\quad x\equiv \frac{r}{R}
\ee
where
\br \label{F-e}
F_\epsilon (x) &=& F^<_\epsilon (x) \equiv \frac{2}{2-\epsilon } x \quad
{\rm for} \quad 0\le x\le 1-\epsilon \nonumber \\
 F_\epsilon (x) &=& F^>_\epsilon (x) \equiv 1-\frac{1}{\epsilon (2-\epsilon )}
(x-1)^2 \quad {\rm for} \quad 1-\epsilon < x \le 1 \nonumber \\
F_\epsilon (x) &=& 1 \quad {\rm for} \quad x>1
\er
and $\epsilon $ is a sufficiently small, positive nonzero parameter. 
It may be checked easily
that $\rho_\epsilon (r)$ is continuous and
has continuous first derivatives everywhere. Further,
the following estimates hold,
\br \label{F-est}
|F_\epsilon (x)| &\le &  k(\epsilon ) x \quad \for \quad 0\le x\le 1
\no \\
|F_\epsilon (x)| &\le &  k(\epsilon ) \quad \for \quad 1 < x<  \infty 
\no \\
|F'_\epsilon (x) | &\le &  k(\epsilon ) \quad \for \quad 0 \le x<  \infty 
\er
where
\be
k(\epsilon) \equiv  \frac{2}{2-\epsilon} >0.
\ee

For the map $\vec N (\vec n)$ with winding number $l$ we choose $l$ 
nonintersecting geodesic discs on the base $S^2$ such
that $\vec N$ varies smoothly from $\vec N = (0,0,1)$ in the center
of each disc to $\vec N = (0,0,-1)$ at the boundary of each disc and 
remains at this value in between the discs. The geodesic radius $\vartheta_0$
of these geodesic discs has to be chosen sufficiently small so that it is
possible to put $l$ nonintersecting discs with this geodesic radius on the
base $S^2$. We prove in Appendix B that the choice
\be \label{vart_0}
\vartheta_0 =\frac{\pi}{4\sqrt{l}} 
\ee
is sufficient. Next we give an explicit expression for the map $\vec N$ 
for one geodesic disc which is chosen symmetric about the north pole.
With 
\be \label{N-chi}
\vec N= (\sin \chi \cos \sigma ,\sin \chi \sin \sigma ,\cos \chi)
\ee
and 
\be
\vec n=(\sin \vartheta \cos \phi ,\sin \vartheta  \sin \phi ,
\cos \vartheta )
\ee
we get
\be
\chi (\vartheta) = \pi F_\epsilon (\tilde\vartheta ) \, , \quad 
\tilde\vartheta \equiv 
\frac{\vartheta}{\vartheta_0}  
\ee
and 
\be
\sigma = \phi
\ee
where $F_\epsilon (\cdot )$ is defined in (\ref{F-e}) (with the additional 
restriction that here $\tilde\vartheta
\le 4\sqrt{l}$ because $\vartheta \le \pi $).
The above-defined $\vec N$ has winding number 1, and to reach winding
number $l$ we have to require that it is nonconstant over $l-1$ 
further geodesic discs. We do not need, however, explicit expressions 
for the contribution of these further discs to $\vec N$, because
their gradients on $S^2$, $|\nabla^{S^2} \vec N|$ 
can be estimated by the same majorants as
the above explicit expression. The reason for this is that the above
symmetrically chosen disc is related to an arbitrary disc with the
same geodesic radius by a rotation of the base space $S^2$, and the
gradient  $|\nabla^{S^2} \vec N|$ is invariant under such rotations.
All that changes is that more regions of the base $S^2$ provide
nonzero contributions to the gradient. 

Next we want to
estimate the energy density ${\cal E}_{\rm Ni} (\vec r) 
\sim |\nabla \vec N |^3$.
Here we first estimate (Einstein summation convention is understood, 
$k,a =1 \ldots 3$)
\be 
|\nabla \vec N|^2 =  (\nabla_k N^a )^2 =
\chi^2_\vartheta (\nabla_k \vartheta )^2 + \sin^2 \chi (\nabla_k \sigma )^2
\ee
where we used (\ref{N-chi}), and $\chi_\vartheta \equiv \partial_\vartheta 
\chi$. The first term may be estimated by
\be
\chi_\vartheta^2 \le \left( \frac{\pi}{\vartheta_0}\right)^2 k(\epsilon)^2
\ee
and, with the help of spherical polar coordinates, by
\br
(\nabla_k \vartheta )^2 &=& \vartheta_r^2 + \frac{1}{r^2} \vartheta_\theta^2 
\no \\
&=& \frac{4\rho_r^2 \cos^2 \rho \sin^2 \theta}{1-\sin^2 \rho \sin^2 \theta}
+ \frac{\sin^2 \rho}{r^2}\frac{4\cos^2 \theta}{1-\sin^2 \rho \sin^2 \theta}
\no \\
&\le& 4\rho_r^2 +4 \frac{\rho^2 }{r^2} \le 8\pi^2 \frac{k(\epsilon)^2}{R^2}
\er
where we used 
\be
\frac{\cos^2 \alpha}{1-\sin^2 \alpha \sin^2 \beta} \le 1 \, ,\quad \sin
\alpha \le \alpha
\ee
as well as Eqs. (\ref{rho-e}) and (\ref{F-est}). For the second term we find
\be
\sin^2 \chi \le \sin^2 \left( \pi k(\epsilon) \frac{\vartheta}{\vartheta_0}
\right) \le
\left( \frac{\pi k(\epsilon)}{\vartheta_0} \right)^2 \sin^2 \vartheta
\ee
where we used 
\be
\sin ax \le a \sin x \quad {\rm for} \quad a\ge 1\, ,\quad x \in 
[0,\frac{\pi}{2}]
\ee
which we prove in Appendix B. Further we find, using $\sigma =\phi$ and 
the expression for $\phi$ in Eq. (\ref{n-rs1})
\be
\sin^2 \vartheta (\nabla_k \phi)^2 = 4\sin^2 \rho \sin^2 \theta (1-
\sin^2 \rho \sin^2 \theta ) \cdot
\no \ee
\be
\cdot \left( \frac{1}{r^2 \sin^2 \theta} + \frac{\rho_r^2 \cos^2 \theta 
+(1/r^2) \cos^2 \rho \sin^2 \rho \sin^2 \theta }{(1 - \sin^2 \rho 
\sin^2 \theta )^2 } \right)
\no \ee
\be
= 4 \frac{\sin^2 \rho}{r^2} (1-\sin^2 \rho \sin^2 \theta ) 
\no \ee
\be
+ 4 \sin^2 \rho \sin^2 \theta \rho_r^2 \frac{\cos^2 \theta}{1 - \sin^2 \rho
\sin^2 \theta} 
\no \ee
\be
+  4 \sin^2 \rho \sin^4 \theta \frac{\sin^2 \rho}{r^2 }
\frac{\cos^2 \rho}{ 1 - \sin^2 \rho \sin^2 \theta }
\no \ee
\be
\le 4 \left( \frac{\pi k(\epsilon)}{R} \right)^2 +
4 \left( \frac{\pi k(\epsilon)}{R} \right)^2 +
4 \left( \frac{\pi k(\epsilon)}{R} \right)^2 =12
 \left( \frac{\pi k(\epsilon)}{R} \right)^2 .
\ee
Putting everything together, we therefore find
\be \label{est-N}
|\nabla \vec N|^2 \le 20 \left( \frac{\pi^2 k(\epsilon)^2}{\vartheta_0 R}
\right)^2 
= 20 l \left( \frac{4\pi k(\epsilon)^2}{R} \right)^2
\ee
Taking the power of $\frac{3}{2}$ and integrating over the ball $B^R$ 
we estimate the energy by
\be
E_{\rm Ni} =\int d^3 r |\nabla \vec N|^3 \le \int_{B^R} d^3 r 5^\frac{3}{2}
l^\frac{3}{2} \left( \frac{8\pi k(\epsilon)^2}{R} \right)^3 
=\frac{4\pi}{3} 5^\frac{3}{2}l^\frac{3}{2} (8\pi k(\epsilon)^2 )^3
\ee
Finally, for the estimate it is sufficient if the estimating field
configuration is a distribution. Therefore we may now perform the
limit $\lim_{\epsilon \to 0} k(\epsilon) =1$ and get, for $Q=l^2$,
\be \label{E-l^2}
E_{\rm Ni} \le   \frac{4\pi}{3} (8\pi )^3   5^\frac{3}{2} 
l^\frac{3}{2} 
\ee

In the cases when the Hopf index $Q$ is not a square, we shall find a
slightly weaker bound (i.e., a slightly larger value for the constant
$C_2$). Again following \cite{LiYa}, we write $Q$ as
\be 
Q= l^2 + m \, ,\quad m \in [1,2l]
\ee
and estimate the energy by the following Hopf map. It maps the ball $B^R$
with radius $R$ about the origin to the two-sphere with a contribution of
$l^2$ to the Hopf index in exactly the way constructed above for the
case $Q=l^2$.  Further, it maps $m$ balls $B^1_i$, $i=1\ldots m$ 
with unit radius to
the $S^2$ such that each ball contributes one unit to the Hopf index. 
Here the $m$ unit balls are chosen such that they intersect neither
each other nor the ball $B^R$. The map from $B^R$ contributes exactly
the above-calculated expression (\ref{E-l^2})
to the energy, whereas the $m$ maps
from the $B^1_i$ contribute the energy of a unit Hopf map each. One might
think to take just the energy (\ref{E-l^2}) with $l=1$ for each energy 
contribution, but we can do better. The reason is that we were forced
to choose $\vartheta_0$ rather small in Eq. (\ref{vart_0}) to ensure that
we can put $l$ geodesic discs on one $S^2$. But here we are only interested in
a map with winding number one, therefore we can choose the ``geodesic
disc'' equal to the full $S^2$, i.e., $\vartheta_0 =\pi$. This leads to
the following energy estimate
\be \label{E-med}
E_{\rm Ni} \le \frac{4\pi}{3} (8\pi )^3 5^\frac{3}{2} l^\frac{3}{2}
+ \frac{4\pi}{3} (2\pi )^3 5^\frac{3}{2} m
\ee
which we want to estimate in terms of $Q$ again, i.e.,
\be
c_2 l^\frac{3}{2} + c_3 m \le c_4 (l^2 +m )^\frac{3}{4} .
\ee
Here the l.h.s. grows linearly in $m$, whereas the r.h.s. grows sublinearly,
therefore the inequality certainly holds for all $m \in [1,2l]$ if it holds
for the maximum value $m=2l$, i.e., 
\be
c_2 l^\frac{3}{2} +2 c_3 l \le c_4 (l^2 +2l )^\frac{3}{4} .
\ee
With the help of the inequality
\be
c_2 + 2 c_3 l^{-\frac{1}{2}} \le c_2 + 2 c_3 \le (c_2 + 2 c_3 )
(1 + \frac{2}{l})^\frac{3}{4}
\ee
we see that the above inequality certainly holds for $c_4 = c_2 + 2c_3$.
Inserting the concrete numbers from (\ref{E-med}) we therefore get
\be
E_{\rm Ni} \le \frac{4\pi}{3} (2\pi)^3
(4^3 + 2 )  5^\frac{3}{2}
 Q^\frac{3}{4}
\ee
as our final estimate for general $Q$.

\subsection{Upper bounds for the AFZ and Faddeev--Niemi models}

We may use our result to obtain upper bounds for the energies
of the AFZ and Faddeev--Niemi models, as well. For the bound on the
AFZ model we use the simple inequality for the energy densities
\be \label{2-4-ineq}
{\cal E}_4 \le \frac{1}{2} {\cal E}_2^2
\ee
 which is obvious from the
$(\vartheta ,\phi)$ parametrization, see Eqs. (\ref{cE-Ni}) and 
(\ref{cE-AFZ}). It follows immediately that $E_{\rm AFZ} \le 
\left( \frac{1}{2} \right)^\frac{3}{4} E_{\rm Ni}$ and, therefore,
\be
E_{\rm AFZ} \le \left( \frac{1}{2} \right)^\frac{3}{4}
\frac{4\pi}{3} (2\pi)^3 (4^3 + 2 )  5^\frac{3}{2}
 Q^\frac{3}{4} .
\ee

For the Faddeev--Niemi model with energy density ${\cal E}_{\rm FN} =
{\cal E}_2 +\lambda {\cal E}_4$ we consider the case $Q=l^2$ of a square Hopf
index first. We use Eq. (\ref{est-N}) for the density
${\cal E}_2$ and the inequality (\ref{2-4-ineq}) and find
\br
E_{\rm FN} & =& \int_{B^R} d^3 r ({\cal E}_2 + \lambda {\cal E}_4 ) \nonumber
\\
& \le &  \int_{B^R} d^3 r \left( 20 l \left( \frac{4\pi}{R} \right)^2
+\frac{\lambda}{2} (20 l)^2 \left( \frac{4\pi}{R} \right)^4 \right) 
\nonumber \\
&=& \frac{4\pi}{3} R^3 \left( 20 l \left( \frac{4\pi}{R} \right)^2
+\frac{\lambda}{2} (20 l)^2 \left( \frac{4\pi}{R} \right)^4 \right) 
\er
and, with the choice 
\be
R=R_0 \sqrt{l}
\ee
we get
\be
E_{\rm FN} \le 20 \frac{4\pi}{3} (4 \pi)^2 l^\frac{3}{2} \left( R_0 
+ 10 (4\pi)^2 \frac{\lambda}{R_0} \right) .
\ee
For a neat estimate we now minimize the expression in brackets w.r.t. 
$R_0$ which leads to
\be
R_0 = 4\pi \sqrt{10 \lambda}
\ee
and to
\be
E_{\rm FN} \le 40 \frac{4\pi}{3} (4\pi)^3 \sqrt{10 \lambda} l^\frac{3}{2} .
\ee
Here we have separated a factor $(4\pi)^3$ because it can be replaced by
$\pi^3$ for winding number one, exactly as in the estimate for the Nicole 
model, which we need for general $Q$. Again we write $Q=l^2 +m$ and estimate
\be
 E_{\rm FN} \le 40 \frac{4\pi}{3} (4\pi)^3 \sqrt{10 \lambda} l^\frac{3}{2}
+ 40 \frac{4\pi}{3} \pi^3 \sqrt{10 \lambda}  m
\ee
and finally by
\be
 E_{\rm FN} \le 40 \frac{4\pi}{3} \pi^3 \sqrt{10 \lambda} (4^3 +2) 
Q^\frac{3}{4}
\ee
in complete analogy to the Nicole model.

\section{Conclusions}

Combining numerical and analytical techniques,
we found the energies for static solutions of the Nicole model within the
symmetric ansatz (\ref{u-mn}) for a wide (in principle arbitrary) range of
the integers $m$ and $n$. It turns out that the energies of these symmetric
solutions grow linearly with the Hopf index $Q=mn$ within a very good
accuracy. Together with the sublinear growth inequality of Section 4,
$E\le C_2 Q^\frac{3}{4}$ this implies that, for sufficiently large values
of $Q$, the symmetric solutions we found cannot be global minima within their
respective topological sectors with fixed Hopf index $Q$. It is of some 
interest to compare this result with the corresponding results for the
AFZ and Faddeev--Niemi models. 

For the AFZ model all solutions for the symmetric ansatz (\ref{u-mn}) and
their corresponding energies can be calculated exactly
\cite{AFZ2}. The energies are
\br \label{sol-AFZ}
E_{\rm AFZ} &=& 8 \sqrt{2} \pi^2 \sqrt{|m||n|(|m| + |n|)} \nonumber \\
&=&    8 \sqrt{2} \pi^2 Q^\frac{3}{4} \sqrt{\mu + \frac{1}{\mu}}
\, ,\quad \mu^2 \equiv \left| \frac{m}{n} \right|
\er
and, for $m=n$ ($\mu =1$) perfectly fit within the upper bound $E_{\rm AFZ} \le
C_2 Q^\frac{3}{4}$ which holds also for the AFZ model (see Section 4).
It is, therefore, plausible to conjecture that these solutions might be
true minima for $m=n$ (and, maybe, even for $m \not= n$ provided that
$m$ and $n$  do not differ too much). 

For the Faddeev--Niemi model the ansatz (\ref{u-mn}) is not compatible with
the equations of motion (due to the lack of conformal symmetry of the
latter), but the cylindrically symmetric ansatz
\be \label{u-m}
u= f(\eta ,\xi) e^{i m \varphi} 
\ee
for complex $f$ is compatible and leads to an equation in the two independent 
variables $\eta$ and $\xi$.  Numerical investigation of the full e.o.m. in
three variables shows that the solutions for the ansatz (\ref{u-m}) are
true minima only for the lowest values $Q=1,2$ of the Hopf index \cite{BS1}. 
In this 
respect, therefore, the Nicole model bears more similarity with the
Faddeev--Niemi model than with the AFZ model. It is tempting to speculate
at this instant that the different behaviour of the AFZ model - i.e. the
possibility that the symmetric solutions (\ref{sol-AFZ}) are true minima
for arbitrary $Q$ - is related to the integrability properties of the latter.
But at the moment this is, of course, only a speculation which deserves
further investigation.   
\\ \\ \\ 
{\large\bf Appendix A: Facts about Hopf maps} \\ \\
A Hopf map is a map from the three-sphere $S^3$ to the two-sphere $S^2$.
The third homotopy group of the two-sphere is nontrivial,
$\Pi_3 (S^2)=\ZZ$, therefore Hopf maps are characterised by an integer 
topological index, the so-called Hopf index $Q$. The three-sphere is 
topologically equivalent to one-point compactified three-dimensional
Euclidean space $\RR^3_0$, therefore, for each Hopf map $S^3 \to S^2$ there 
exists a corresponding map $\RR^3_0 \to S^2$, which we shall call Hopf map,
as well, and which has the same Hopf index (due to the metric independence of
a topological index). Explicitly, a Hopf map $\RR^3_0 \to S^2$ may be
given by a three-component unit vector field
\be
\vec n (\vec r) \, : \, \RR^3_0 \to S^2 \, ,\quad 
\vec n^2 =1 \, ,\quad \lim_{\vec r\to \infty}
=\vec n_0 = {\rm const}
\ee
where the tip of the unit vector field spans the unit two-sphere, or via
stereographic projection
\be
{\vec n} = {1\o {1+\mid u\mid^2}} \, ( u+\bar u , -i ( u-\bar u ) ,  
1-u\bar u ) \; ;
\qquad
u  = \frac{n_1 + i n_2}{1 + n_3}.
\label{stereo}
\ee
by a complex field
\be
u(\vec r) \, : \, \RR^3_0 \to \CC_0 \, , \quad \lim_{\vec r\to \infty}
= u_0 = {\rm const}
\ee
where our conventions are such that the projection is from the south pole to 
the equatorial plane of the two-sphere. Further, $\CC_0$ is the one-point
compactified complex plane. A third possibility to parametrize a Hopf map 
is by the spherical angles $\phi \in [0,2\pi ]$, $\vartheta \in [0,\pi ]$,
which are related to the unit vector $\vec n$ and complex field $u$ via
\be
u= \tan \frac{\vartheta}{2} e^{i\phi} \, ,\quad 
\vec n = ( \sin \vartheta \cos \phi , \sin \vartheta \sin \phi ,\cos \vartheta
) .
\ee

The geometry behind the Hopf map may be visualised as follows. The pre-images
under the inverse of the Hopf map $u$ of points in the target space $S^2$ (or, 
equivalently, $\CC_0$) are closed curves in $\RR^3_0$ (in general, knots), 
where each two curves corresponding 
to two different points in target space are linked exactly $Q$ times.
An analytic expression for the Hopf index $Q$ is 
\be
Q=\frac{1}{16 \pi^2}\int d^3 r \vec {\cal A} \cdot \vec {\cal B}
\ee
where $\vec {\cal B}$ is the Hopf curvature
\be \label{ho-curv}
\vec {\cal B} =\frac{2}{i} \frac{\nabla u \times \nabla 
\bar u}{(1 + u\bar u)^2} = -\frac{1}{2} \epsilon^{abc} n^a \nabla n^b \times
\nabla n^c = -\sin \vartheta \nabla \vartheta \times \nabla \phi
\ee
and $\vec {\cal A}$ is the gauge potential for the ``magnetic field''
$\vec{\cal B}$, $\vec {\cal B} = \nabla \times \vec {\cal A}$.  
There is no local expression for $\vec {\cal A}$ in terms of the Hopf map
$u$ alone (however, there certainly exist  non-local expressions for certain 
gauges, as is obvious from the analogy with electrodynamics, by choosing, 
e.g., the Couloumb gauge $\nabla \cdot \vec {\cal A}=0$ which implies
$\vec {\cal A}= \nabla \times \int d^3 r' \Delta^{-1} (\vec r -\vec r')
\vec {\cal B}(\vec r')$).   

Geometrically, $\vec{\cal B}$ is the Hodge dual of the pullback of the
area two-form on $S^2$ under the Hopf map $u$, i.e., 
\be
{\cal B} \equiv {\cal B}_k dr^k =*{\cal F} ,
\ee
\be
{\cal F} \equiv \frac{1}{2}
{\cal F}_{jk}dr^j dr^k \, ,\quad 
{\cal B}_k = \varepsilon_{klm}{\cal F}_{lm}
\ee
where
\be
{\cal F}=u^* (\Omega) \, , \quad \Omega =\frac{2}{i}\frac{d\zeta d\bar \zeta}{
(1+\zeta \bar \zeta)^2} \, ,\quad \int_{S^2}\Omega =4\pi .
\ee
The second cohomology group of the three-sphere is trivial, $H_2 (S^3)=0$,
therefore the closed two-form ${\cal F}$ must be exact, ${\cal F}=d{\cal A}$.
It follows that the gauge potential $\vec {\cal A}$ is globally well-defined
for appropriately chosen gauges (e.g., in the Couloumb gauge).

The simplest Hopf map with Hopf index 1 is
\be \label{s-ho1}
u =i\frac{2(x +iy)}{2z +i(r^2 -1)} 
\ee
(the irrelevant factor $i$ has been introduced to be in exact agreement with
the simplest Hopf map of Eq. (\ref{s-ho})). Its level curves (i.e., the
pre-images of points $u= {\rm const}$) are circles which lie on tori, and each
two different circles are linked precisely once. 

There are different ways to construct Hopf maps explicitly. One method which
we shall need is to compose a given Hopf map with a map $S^2 \to S^2$,
\be
\vec N (\vec n(\vec r))\, : \, \RR^3_0 \stackrel{\vec n(\vec r)}{\to}
S^2 \stackrel{\vec N(\vec n)}{\to} S^2
\ee
If the Hopf map $\vec n(\vec r)$ has Hopf index $Q$ and the map 
$\vec N(\vec n)$ has winding number $w$, then the composition map
$\vec N(\vec n(\vec r))$ has Hopf index $Q' = w^2 Q$.

A method to construct all possible Hopf maps starts from maps $S^3 \to S^3$
(or, equivalently, maps $\RR^3_0 \to S^3$). The third homotopy group
of $S^3$ is nontrivial, $\Pi_3 (S^3) =\ZZ$, therefore such maps are
classified by a topological index, the winding number $W$. It is possible
to construct Hopf maps $\RR^3_0 \to S^2$ from maps $\RR^3_0 \to S^3$  
such that the Hopf index equals the winding number.
Explicitly a map $\RR^3_0 \to S^3$ may be given by a four component unit
vector field 
\be
( e_1 (\vec r) ,e_2 (\vec r) ,e_3 (\vec r) ,e_4 (\vec r)) \, : \,
\RR^3_0 \to S^3 \, ,\quad e_\alpha e_\alpha =1 \, ,\quad \alpha =1 ,\ldots ,4
\nonumber
\ee
\be
\lim_{\vec r\to \infty }e_\alpha =e_\alpha^0 = {\rm const}
\ee
then the corresponding Hopf map is given in terms of $u$ as
\be
u=\frac{e_1 +ie_2}{e_3 +ie_4}
\ee
or in terms of $\vec n$ as
\be \label{n-s3}
\vec n = (2e_1 e_3 + 2e_2 e_4 ,-2e_1 e_4 +2e_2 e_3, 
e_3^2 +e_4^2  -e_1^2 -e_2^2  ).
\ee
Further, it is now possible to give an explicit, local expression for the gauge
potential $\vec {\cal A}$, in terms of the $e_\alpha$, as
\be
\vec {\cal A} = \frac{2}{i} [(e_1 -ie_2 ) \nabla (e_1 +ie_2) +
(e_3 -ie_4 )\nabla (e_3 +ie_4 )] .
\ee
It may be checked without difficulty that indeed $\nabla \times \vec {\cal A}
=\vec {\cal B}$ where  $\vec {\cal B}$ is the Hopf curvature (\ref{ho-curv}).

As the group manifold of $SU(2)$ is equivalent to the three-sphere $S^3$,
we may use maps $\RR_0^3 \to SU(2)$ instead. Indeed, with the group 
element
\be
U=e_4 -i\vec e \cdot \vec \sigma \equiv \exp [-i\rho \vec k \cdot \vec \sigma ]
\, ,\quad \lim_{\vec r\to \infty} U=U_0 ={\rm const}
\ee
\be
e_4 =\cos \rho \, ,\quad \vec e =\sin \rho \, \vec k \, ,\quad \vec k^2 =1
\ee
(where $\vec e = (e_1 ,e_2 ,e_3)$ and $\vec \sigma$ are the Pauli matrices)
the Hopf map (\ref{n-s3}) may be expressed like
\be
\vec n = \frac{1}{2} {\rm tr} \, \sigma _3 U^\dagger \vec \sigma U.
\ee
This latter representation may be used to produce ansaetze for Hopf maps
with cylindrical symmetry from maps $\RR_0^3 \to SU(2)$ with rotational
symmetry. Indeed, for 
\be \label{q-s3-r}
\vec k =\hat r \equiv \vec r / r \, ,\quad \rho = \rho (r) \, ,\quad
\rho (0) =0 \, ,\quad \lim_{r\to \infty } \rho =\pi W \, ,\quad
W\in\ZZ 
\ee
(the conditions on $\rho (r)$ are to ensure a well-defined $U$ on all
$\RR_0^3$, and they provide, at the same time, a winding number equal to $W$
for the corresponding map $\RR_0^3 \to S^3$ or $\RR_0^3 \to SU(2)$),
the resulting group element $U(\rho (r),\hat r )$ is rotationally invariant
in the sense that any rotation in $\RR_0^3$ can be compensated by a
$SU(2)$ transformation,
\be
U(\rho (r), O\hat r) \equiv \cos \rho(r) -i \sin (\rho (r) \vec \sigma
\cdot (O\hat r) =
V^\dagger U(\rho (r), \hat r) V
\ee
where $O$ is a rotation matrix acting on $\hat r$ and $V$ is the $SU(2)$
matrix corresponding to the rotation $O$. 
Within this ansatz, and using spherical polar coordinates $\hat r =
(\sin \theta \cos \varphi ,\sin \theta \sin \varphi ,\cos \theta)$, we get
for the complex field $u$
\be
u=\frac{\sin \rho \sin \theta e^{i\varphi}}{\sin \rho \cos \theta +
i\cos \rho} 
\ee
which has cylindrical symmetry in the sense that any rotation $\varphi
\to \varphi + \varphi_0$ can be compensated by a phase transformation 
$u\to e^{i\varphi_0} u$. Further, $u$ is a Hopf map with Hopf index equal
to the winding number $W$.
For technical reasons we prefer to use a $u$ which 
is multiplied by $-i$ in our main calculation, i.e.,
\be
u =-i\frac{e_1 +i e_2}{e_3 +i e_4} = \frac{\sin \rho \sin \theta}{
\sqrt{\sin^2 \rho \cos^2 \theta +\cos^2 \rho}}e^{i\varphi +i \arctan[ \tan \rho
\cos \theta]} .
\ee
With $u=\tan (\vartheta /2) e^{i\phi}$ this leads to
\be \label{n-rs1}
\sin \vartheta = 2\sin \rho \sin \theta \sqrt{1 - \sin^2 \rho \sin^2 \theta}
\, ,\quad \phi = \varphi + \arctan [\tan\rho \cos \theta ].
\ee 
Finally, we display the energy densities of the Nicole and AFZ models in the 
three different parametrizations, $ {\cal E}_{\rm Ni}={\cal E}_2^\frac{3}{2}$ 
with
\be \label{cE-Ni}
{\cal E}_2 =
|\nabla \vec n|^2 \equiv |\nabla_k n^a |^2 = 4\frac{\nabla u \cdot
\nabla \bar u}{(1+u\bar u)^2} = (\nabla \vartheta )^2 + \sin^2 \vartheta
(\nabla \phi )^2 
\ee
and ${\cal E}_{\rm AFZ} = {\cal E}_4^\frac{3}{4} $ with
\bdi
{\cal E}_4 = \vec {\cal B}^2 =\frac{1}{2} \left( (\nabla n^b \cdot \nabla n^b 
)^2 - (\nabla n^b \cdot \nabla n^c )^2 \right) =
\edi
\be \label{cE-AFZ}
4\frac{(\nabla u \cdot \nabla \bar u)^2 - (\nabla u)^2 (\nabla \bar u )^2 }{
(1 + u\bar u )^4} = \sin^2 \vartheta \left( (\nabla \vartheta )^2 
(\nabla \phi )^2 - (\nabla \vartheta \cdot \nabla \phi )^2 \right) .
\ee
\\ \\ \\
{\large\bf Appendix B} \\ \\
{\bf B1: Proof that 
\be \label{proofB1}
(a+b)^\frac{3}{2} \ge c\, a^\frac{1}{2} b \qquad {\rm for} 
\quad a,b \ge 0\, ,\quad
c \le \sqrt{3} \frac{3}{2} .
\ee
} \\ \\
By squaring the above expression and subtracting the r.h.s. we get
\be
a^3 + 3 a^2 b + 3(1-d^2) ab^2 + b^3 \ge 0 \, ,\qquad d^2 \equiv \frac{c^2}{3}
\ee
or, by introducing the variable $x=\frac{a}{b}$,
\be
h(x) \equiv x^3 + 3 x^2 + 3(1-d^2) x +1 \ge 0 \qquad {\rm for} \qquad x\ge 0.
\ee
This problem we solve by first calculating the minimum $x_{\rm min}$
of $h(x)$ for $x\ge 0$, and by then requiring that $h(x_{\rm min})=0$.
The minimum is at
\be
h'(x) = 3 (x^2 + 2x + (1-d^2)) \equiv 0 \quad \Rightarrow \quad x_{\rm min}
=d-1
\ee
and the condition $h(x_{\rm min})=0$ finally leads to
\be
h(x_{\rm min}) = d^2 (3-2d) \equiv 0 \quad \wedge \quad d \ge 1 
\qquad \Rightarrow \qquad d=\frac{3}{2}
\ee
which implies  (\ref{proofB1}). \hfill QED  \\ \\
{\bf B2: Proof that one may cover the unit sphere with $l$ nonintersecting 
discs with geodesic radius $\vartheta_0 = \frac{\pi}{4\sqrt{l}}$.} \\ \\
First we study the case where $l$ is a square, $l=k^2$. We divide the
northern hemisphere of the two-sphere into segments by circles which are
concentric about the north pole and are given by
\be
\vartheta = \frac{m\pi}{2k} \, ,\quad m=1 \ldots k .
\ee
This divides the northern hemisphere into one disc with geodesic radius
$\frac{\pi}{2k}$ about the north pole (on which one geodesic disc with
geodesic radius $\vartheta_0 = \frac{\pi}{4k}$ certainly does fit) and
into sphere segments such that the boundary circles have a geodesic 
distance of $\frac{\pi}{2k}$. Therefore discs with geodesic diameter
$2\vartheta_0 =\frac{\pi}{2k}$ exactly fit in between. Further, $2m+1$ 
geodesic discs fit on the segment between the circles at 
$ \vartheta = \frac{m\pi}{2k}$ and at $\vartheta = \frac{(m+1)\pi}{2k}$. 
This is because the central circle of the segment has a circumference 
wich is bigger
than the sum of $m+1$ diameters of the geodesic discs, i.e., 
\be
2\pi \sin \left( (m+ \frac{1}{2}) \frac{\pi}{2k} \right) \ge
2\pi \frac{2}{\pi} \left( (m+ \frac{1}{2}) \frac{\pi}{2k} \right) =
(2m +1 ) \frac{\pi}{k} \ge (2m +1 ) \frac{\pi}{2k}
\ee
where we used
\be
\sin x \ge \frac{2}{\pi} x \quad {\rm for} \quad x \in [0,\frac{\pi}{2}] .
\ee
Finally, summing over all discs on all segments (including the first disc at
the segment which is itself a disc) we find
\be
\sum_{m=0}^{k-1} (2m+1) = \sum_{m=1}^k (2m-1) =k^2
\ee
therefore we may indeed distribute $k^2$ nonintersecting geodesic discs
with geodesic radius $\vartheta_0 =\frac{\pi}{4k}$ on the northern
hemisphere. 

It remains to study the case where $l$ is not a square. Then we write
$l=k^2 +j$. We distribute the $k^2$ discs on the northern hemisphere as just
described. Further, we distribute the remaining $j$ discs on the
southern hemisphere. This is obviously possible as long as
$k^2 \ge j$. So we cover all cases except the case $l=3$. This latter
case may be proven easily with the help of elementary geometry.
\hfill QED  \\ \\
{\bf B3: Proof that 
\be
\sin ax \le a \sin x \quad {\rm for} \quad
a\ge 1 \, , \quad x\in [0,\frac{\pi}{2}]
\ee} \\ \\
We introduce $a\equiv 1+\epsilon$, $\epsilon \ge 0$ and write
$$
\sin \left( (1+\epsilon )x \right) = \sin x \cos \epsilon x +
\cos x \sin \epsilon x  
$$
\be
\le \sin x + \epsilon x \cos x \le 
\sin x + \epsilon \sin x \equiv a \sin x
\ee
where we used $\cos x \le 1$, $ \sin x \le x$ for $x\ge 0$, and 
\be
x\cos x \le \sin x \quad \Rightarrow \quad \tan x \ge x \quad
 {\rm for} \quad x\in [0,\frac{\pi}{2}]
\ee
which is obviously true. \hfill QED
\\ \\ \\
{\large\bf Appendix C: \\ On the possibility of
a lower bound $C_1 Q^\frac{3}{4} \le E_Q$}
\\ \\
It may be useful to briefly comment on the problem of deriving a lower bound
for the energy of the type $C_1 Q^\frac{3}{4} \le E_Q$, analogously to the 
case of the Faddeev--Niemi model, where this bound can be proven
(Vakulenko--Kapitansky bound, see e.g. \cite{VaKa}, \cite{Ward1}, 
\cite{Shab1}).
In the case of the Faddeev--Niemi model,
the first step in the proof is  Hoelder's inequality (we use the 
conventions of \cite{Evans}, where all inequalities of this appendix can
be found; further, we assume that $\vec {\cal A}$ and $\vec {\cal B}$ belong
to the appropriate Sobolev spaces such that all integrals below exist) 
\be \label{hoeld}
16 \pi^2 Q \equiv
\int d^3 r \vec {\cal A}\cdot \vec {\cal B} \le \left( \int d^3 r 
|\vec {\cal A} |^p 
\right)^\frac{1}{p} \left( \int d^3 r |\vec {\cal B}|^q \right)^\frac{1}{q}
\ee
\be
\frac{1}{p} + \frac{1}{q} =1.
\ee
For the 
Faddeev--Niemi model one has to choose $p=6$, $q=\frac{6}{5}$ and may 
estimate the first term with the help of the Gagliardo--Nirenberg--Sobolev
(GNS) inequality,
\be
\left( \int d^3 r |\vec {\cal A}|^6 \right)^\frac{1}{6} \le
c \left( \int d^3 r |\nabla |\vec {\cal A}||^2 \right)^\frac{1}{2}
\ee
(here and below $c$ is an unspecified constant)
and, when the Couloumb gauge condition $\nabla \cdot \vec {\cal A}=0$ 
is imposed,
the integrand can be reexpressed like
\be
|\nabla |\vec {\cal A}||^2 \le c |\sum_{jk} \partial_j {\cal A}_k |^2 
= c (|\nabla \times
\vec {\cal A}|^2 + \nabla \cdot \vec \Lambda )
\ee
where 
\be
\Lambda_1 = ({\cal A}_2 \partial_2 +{\cal A}_3 \partial_3 ) {\cal A}_1 - 
{\cal A}_1 (\partial_2 {\cal A}_2 +  \partial_3 {\cal A}_3 )
\ee
and $\Lambda_2$, $\Lambda_3$ follow by permutation.
Therefore, the integrand is the sum of $\vec {\cal B}^2$ and a total 
divergence which does
not contribute to the integral (for $\vec {\cal A}$ which decay sufficiently
fast in the limit $|\vec r | \to \infty$; this is automatically satisfied
by the class of $\vec {\cal A}$ we consider). 
The proof continues with the interpolation inequality
\be
\left( \int d^3 r |\vec {\cal B}|^\frac{6}{5} \right)^\frac{5}{6}
\le \left( \int d^3 r |\vec {\cal B}| \right)^\frac{2}{3}
\left( \int d^3 r |\vec {\cal B}|^2 \right)^\frac{1}{6}
\ee
for the second term. This leads to
\be
16 \pi^2 Q \le c 
\left( \int d^3 r |\vec {\cal B}|^2 \right)^\frac{2}{3}
\left( \int d^3 r |\vec {\cal B}| \right)^\frac{2}{3}
\ee
and, together with $|\vec {\cal B}|^2 ={\cal E}_4$ and $|\vec {\cal B}|
\le \frac{1}{\sqrt{2}} {\cal E}_2$, to
\br
16 \pi^2 Q & \le & c \lambda^{-\frac{2}{3}} \left[ \left( \lambda \int
d^3 r {\cal E}_4 \right)^\frac{1}{2} \left( \int d^3 r {\cal E}_2 
\right)^\frac{1}{2} \right]^\frac{4}{3} \nonumber \\
& \le & c\lambda^{-\frac{2}{3}} \left[ \lambda \int d^3 r {\cal E}_4 
+ \int d^3 r {\cal E}_2 \right]^\frac{4}{3}
\er
where we used Cauchy's inequality in the last step. The 
Vakulenko--Kapitansky bound follows immediately by taking the above
inequality to the power $\frac{3}{4}$. 

Now let us point out where the analogous proof fails for the
AFZ and Nicole models. One again starts with Hoelder's inequality
(\ref{hoeld}), but now one would have to choose $p=3$ and
$ q=\frac{3}{2}$. With the GNS inequality one can again estimate the first
term 
\be
\left( \int d^3 r |\vec {\cal A}|^3 \right)^\frac{1}{3} \le
c \left( \int d^3 r |\nabla |\vec {\cal A}||^\frac{3}{2} \right)^\frac{2}{3}
\ee
but now the integrand is
\be
|\nabla |\vec {\cal A}||^\frac{3}{2} \le c |\sum_{jk} \partial_j {\cal A}_k 
|^\frac{3}{2} 
= c \left( |\nabla \times
\vec {\cal A}|^2 + \nabla \cdot \vec \Lambda \right)^\frac{3}{4}
\ee
and therefore no longer a total derivative, and we cannot express the 
integrand in terms of the Hopf curvature alone. 
We have not been able to overcome this difficulty.
For our main results, however, the upper bound is more important, which
shows that for sufficiently high Hopf index our solutions of Section 2
cannot be true minimizers of the energy.
\\ \\ \\
{\large\bf Acknowledgement:} \\
This research was partly supported by MCyT(Spain) and FEDER
(FPA2005-01963), Incentivos from Xunta de Galicia and the EC
network "EUCLID". Further, 
we thank "Centro de computacion de Galicia" (CESGA) for computer support.
CA acknowledges support from the
Austrian START award project FWF-Y-137-TEC and from the  FWF
project P161 05 NO 5 of N.J. Mauser. AW gratefully acknowledges
support from the Polish Ministry of Education and Science.

\newpage 

\def\temp{1.34}%
\let\tempp=\relax
\expandafter\ifx\csname psboxversion\endcsname\relax
  \message{PSBOX(\temp) loading}%
\else
    \ifdim\temp cm>\psboxversion cm
      \message{PSBOX(\temp) loading}%
    \else
      \message{PSBOX(\psboxversion) is already loaded: I won't load
        PSBOX(\temp)!}%
      \let\temp=\psboxversion
      \let\tempp= 
    \fi
\fi
\tempp
\let\psboxversion=\temp
\catcode`\@=11
%
%
\def\psfortextures{
\def\PSspeci@l##1##2{%
\special{illustration ##1\space scaled ##2}%
}}%
\def\psfordvitops{
\def\PSspeci@l##1##2{%
\special{dvitops: import ##1\space \the\drawingwd \the\drawinght}%
}}%
\def\psfordvips{
\def\PSspeci@l##1##2{%
\d@my=0.1bp \d@mx=\drawingwd \divide\d@mx by\d@my
\includegraphics{##1\space}}}%
\def\psforoztex{
\def\PSspeci@l##1##2{%
\special{##1 \space
      ##2 1000 div dup scale
      \number-\psllx\space \number-\pslly\space translate
}}}%
\def\psfordvitps{
\def\psdimt@n@sp##1{\d@mx=##1\relax\edef\psn@sp{\number\d@mx}}
\def\PSspeci@l##1##2{%
\special{dvitps: Include0 "psfig.psr"}
\psdimt@n@sp{\drawingwd}
\special{dvitps: Literal "\psn@sp\space"}
\psdimt@n@sp{\drawinght}
\special{dvitps: Literal "\psn@sp\space"}
\psdimt@n@sp{\psllx bp}
\special{dvitps: Literal "\psn@sp\space"}
\psdimt@n@sp{\pslly bp}
\special{dvitps: Literal "\psn@sp\space"}
\psdimt@n@sp{\psurx bp}
\special{dvitps: Literal "\psn@sp\space"}
\psdimt@n@sp{\psury bp}
\special{dvitps: Literal "\psn@sp\space startTexFig\space"}
\special{dvitps: Include1 "##1"}
\special{dvitps: Literal "endTexFig\space"}
}}%
\def\psfordvialw{
\def\PSspeci@l##1##2{
\special{language "PostScript",
position = "bottom left",
literal "  \psllx\space \pslly\space translate
  ##2 1000 div dup scale
  -\psllx\space -\pslly\space translate",
include "##1"}
}}%
\def\psforptips{
\def\PSspeci@l##1##2{{
\d@mx=\psurx bp
\advance \d@mx by -\psllx bp
\divide \d@mx by 1000\multiply\d@mx by \xscale
\incm{\d@mx}
\let\tmpx\dimincm
\d@my=\psury bp
\advance \d@my by -\pslly bp
\divide \d@my by 1000\multiply\d@my by \xscale
\incm{\d@my}
\let\tmpy\dimincm
\d@mx=-\psllx bp
\divide \d@mx by 1000\multiply\d@mx by \xscale
\d@my=-\pslly bp
\divide \d@my by 1000\multiply\d@my by \xscale
\at(\d@mx;\d@my){\special{ps:##1 x=\tmpx, y=\tmpy}}
}}}%
\def\psonlyboxes{
\def\PSspeci@l##1##2{%
\at(0cm;0cm){\boxit{\vbox to\drawinght
  {\vss\hbox to\drawingwd{\at(0cm;0cm){\hbox{({\tt##1})}}\hss}}}}
}}%
\def\psloc@lerr#1{%
\let\savedPSspeci@l=\PSspeci@l%
\def\PSspeci@l##1##2{%
\at(0cm;0cm){\boxit{\vbox to\drawinght
  {\vss\hbox to\drawingwd{\at(0cm;0cm){\hbox{({\tt##1}) #1}}\hss}}}}
\let\PSspeci@l=\savedPSspeci@l
}}%
%
%
\newread\pst@mpin
\newdimen\drawinght\newdimen\drawingwd
\newdimen\psxoffset\newdimen\psyoffset
\newbox\drawingBox
\newcount\xscale \newcount\yscale \newdimen\pscm\pscm=1cm
\newdimen\d@mx \newdimen\d@my
\newdimen\pswdincr \newdimen\pshtincr
\let\ps@nnotation=\relax
{\catcode`\|=0 |catcode`|\=12 |catcode`|
|catcode`#=12 |catcode`*=14
|xdef|backslashother{\}*
|xdef|percentother{
|xdef|tildeother{~}*
|xdef|sharpother{#}*
}%
\def\R@moveMeaningHeader#1:->{}%
\def\uncatcode#1{%
\edef#1{\expandafter\R@moveMeaningHeader\meaning#1}}%
\def\execute#1{#1}
\def\psm@keother#1{\catcode`#112\relax}
\def\executeinspecs#1{%
\execute{\begingroup\let\do\psm@keother\dospecials\catcode`\^^M=9#1\endgroup}}%
\def\@mpty{}%
\def\matchexpin#1#2{
  \fi%
  \edef\tmpb{{#2}}%
  \expandafter\makem@tchtmp\tmpb%
  \edef\tmpa{#1}\edef\tmpb{#2}%
  \expandafter\expandafter\expandafter\m@tchtmp\expandafter\tmpa\tmpb\endm@tch%
  \if\match%
}%
\def\matchin#1#2{%
  \fi%
  \makem@tchtmp{#2}%
  \m@tchtmp#1#2\endm@tch%
  \if\match%
}%
\def\makem@tchtmp#1{\def\m@tchtmp##1#1##2\endm@tch{%
  \def\tmpa{##1}\def\tmpb{##2}\let\m@tchtmp=\relax%
  \ifx\tmpb\@mpty\def\match{YN}%
  \else\def\match{YY}\fi%
}}%
\def\incm#1{{\psxoffset=1cm\d@my=#1
 \d@mx=\d@my
  \divide\d@mx by \psxoffset
  \xdef\dimincm{\number\d@mx.}
  \advance\d@my by -\number\d@mx cm
  \multiply\d@my by 100
 \d@mx=\d@my
  \divide\d@mx by \psxoffset
  \edef\dimincm{\dimincm\number\d@mx}
  \advance\d@my by -\number\d@mx cm
  \multiply\d@my by 100
 \d@mx=\d@my
  \divide\d@mx by \psxoffset
  \xdef\dimincm{\dimincm\number\d@mx}
}}%
%
\newif\ifNotB@undingBox
\newhelp\PShelp{Proceed: you'll have a 5cm square blank box instead of
your graphics (Jean Orloff).}%
\def\s@tsize#1 #2 #3 #4\@ndsize{
  \def\psllx{#1}\def\pslly{#2}%
  \def\psurx{#3}\def\psury{#4}
  \ifx\psurx\@mpty\NotB@undingBoxtrue
  \else
    \drawinght=#4bp\advance\drawinght by-#2bp
    \drawingwd=#3bp\advance\drawingwd by-#1bp
  \fi
  }%
\def\sc@nBBline#1:#2\@ndBBline{\edef\p@rameter{#1}\edef\v@lue{#2}}%
\def\g@bblefirstblank#1#2:{\ifx#1 \else#1\fi#2}%
{\catcode`\%=12
\xdef\B@undingBox{
\def\ReadPSize#1{
 \readfilename#1\relax
 \let\PSfilename=\lastreadfilename
 \openin\pst@mpin=#1\relax
 \ifeof\pst@mpin \errhelp=\PShelp
   \errmessage{I haven't found your postscript file (\PSfilename)}%
   \psloc@lerr{was not found}%
   \s@tsize 0 0 142 142\@ndsize
   \closein\pst@mpin
 \else
   \if\matchexpin{\GlobalInputList}{, \lastreadfilename}%
   \else\xdef\GlobalInputList{\GlobalInputList, \lastreadfilename}%
     \immediate\write\psbj@inaux{\lastreadfilename,}%
   \fi%
   \loop
     \executeinspecs{\catcode`\ =10\global\read\pst@mpin to\n@xtline}%
     \ifeof\pst@mpin
       \errhelp=\PShelp
       \errmessage{(\PSfilename) is not an Encapsulated PostScript File:
           I could not find any \B@undingBox: line.}%
       \edef\v@lue{0 0 142 142:}%
       \psloc@lerr{is not an EPSFile}%
       \NotB@undingBoxfalse
     \else
       \expandafter\sc@nBBline\n@xtline:\@ndBBline
       \ifx\p@rameter\B@undingBox\NotB@undingBoxfalse
         \edef\t@mp{%
           \expandafter\g@bblefirstblank\v@lue\space\space\space}%
         \expandafter\s@tsize\t@mp\@ndsize
       \else\NotB@undingBoxtrue
       \fi
     \fi
   \ifNotB@undingBox\repeat
   \closein\pst@mpin
 \fi
\message{#1}%
}%
%
%
\def\psboxto(#1;#2)#3{\vbox{%
   \ReadPSize{#3}%
   \advance\pswdincr by \drawingwd
   \advance\pshtincr by \drawinght
   \divide\pswdincr by 1000
   \divide\pshtincr by 1000
   \d@mx=#1
   \ifdim\d@mx=0pt\xscale=1000
         \else \xscale=\d@mx \divide \xscale by \pswdincr\fi
   \d@my=#2
   \ifdim\d@my=0pt\yscale=1000
         \else \yscale=\d@my \divide \yscale by \pshtincr\fi
   \ifnum\yscale=1000
         \else\ifnum\xscale=1000\xscale=\yscale
                    \else\ifnum\yscale<\xscale\xscale=\yscale\fi
              \fi
   \fi
   \divide\drawingwd by1000 \multiply\drawingwd by\xscale
   \divide\drawinght by1000 \multiply\drawinght by\xscale
   \divide\psxoffset by1000 \multiply\psxoffset by\xscale
   \divide\psyoffset by1000 \multiply\psyoffset by\xscale
   \global\divide\pscm by 1000
   \global\multiply\pscm by\xscale
   \multiply\pswdincr by\xscale \multiply\pshtincr by\xscale
   \ifdim\d@mx=0pt\d@mx=\pswdincr\fi
   \ifdim\d@my=0pt\d@my=\pshtincr\fi
   \message{scaled \the\xscale}%
 \hbox to\d@mx{\hss\vbox to\d@my{\vss
   \global\setbox\drawingBox=\hbox to 0pt{\kern\psxoffset\vbox to 0pt{%
      \kern-\psyoffset
      \PSspeci@l{\PSfilename}{\the\xscale}%
      \vss}\hss\ps@nnotation}%
   \global\wd\drawingBox=\the\pswdincr
   \global\ht\drawingBox=\the\pshtincr
   \global\drawingwd=\pswdincr
   \global\drawinght=\pshtincr
   \baselineskip=0pt
   \copy\drawingBox
 \vss}\hss}%
  \global\psxoffset=0pt
  \global\psyoffset=0pt
  \global\pswdincr=0pt
  \global\pshtincr=0pt 
  \global\pscm=1cm 
}}%
%
%
\def\psboxscaled#1#2{\vbox{%
  \ReadPSize{#2}%
  \xscale=#1
  \message{scaled \the\xscale}%
  \divide\pswdincr by 1000 \multiply\pswdincr by \xscale
  \divide\pshtincr by 1000 \multiply\pshtincr by \xscale
  \divide\psxoffset by1000 \multiply\psxoffset by\xscale
  \divide\psyoffset by1000 \multiply\psyoffset by\xscale
  \divide\drawingwd by1000 \multiply\drawingwd by\xscale
  \divide\drawinght by1000 \multiply\drawinght by\xscale
  \global\divide\pscm by 1000
  \global\multiply\pscm by\xscale
  \global\setbox\drawingBox=\hbox to 0pt{\kern\psxoffset\vbox to 0pt{%
     \kern-\psyoffset
     \PSspeci@l{\PSfilename}{\the\xscale}%
     \vss}\hss\ps@nnotation}%
  \advance\pswdincr by \drawingwd
  \advance\pshtincr by \drawinght
  \global\wd\drawingBox=\the\pswdincr
  \global\ht\drawingBox=\the\pshtincr
  \global\drawingwd=\pswdincr
  \global\drawinght=\pshtincr
  \baselineskip=0pt
  \copy\drawingBox
  \global\psxoffset=0pt
  \global\psyoffset=0pt
  \global\pswdincr=0pt
  \global\pshtincr=0pt 
  \global\pscm=1cm
}}%
%
\def\psbox#1{\psboxscaled{1000}{#1}}%
\newif\ifn@teof\n@teoftrue
\newif\ifc@ntrolline
\newif\ifmatch
\newread\j@insplitin
\newwrite\j@insplitout
\newwrite\psbj@inaux
\immediate\openout\psbj@inaux=psbjoin.aux
\immediate\write\psbj@inaux{\string\joinfiles}%
\immediate\write\psbj@inaux{\jobname,}%
%
%
\def\toother#1{\ifcat\relax#1\else\expandafter%
  \toother@ux\meaning#1\endtoother@ux\fi}%
\def\toother@ux#1 #2#3\endtoother@ux{\def\tmp{#3}%
  \ifx\tmp\@mpty\def\tmp{#2}\let\next=\relax%
  \else\def\next{\toother@ux#2#3\endtoother@ux}\fi%
\next}%
%
%
\let\readfilenamehook=\relax
\def\re@d{\expandafter\re@daux}
\def\re@daux{\futurelet\nextchar\stopre@dtest}%
\def\re@dnext{\xdef\lastreadfilename{\lastreadfilename\nextchar}%
  \afterassignment\re@d\let\nextchar}%
\def\stopre@d{\egroup\readfilenamehook}%
\def\stopre@dtest{%
  \ifcat\nextchar\relax\let\nextread\stopre@d
  \else
    \ifcat\nextchar\space\def\nextread{%
      \afterassignment\stopre@d\chardef\nextchar=`}%
    \else\let\nextread=\re@dnext
      \toother\nextchar
      \edef\nextchar{\tmp}%
    \fi
  \fi\nextread}%
\def\readfilename{\bgroup%
  \let\\=\backslashother \let\%=\percentother \let\~=\tildeother
  \let\#=\sharpother \xdef\lastreadfilename{}%
  \re@d}%
%
%
\xdef\GlobalInputList{\jobname}%
\def\psnewinput{%
  \def\readfilenamehook{
    \if\matchexpin{\GlobalInputList}{, \lastreadfilename}%
    \else\xdef\GlobalInputList{\GlobalInputList, \lastreadfilename}%
      \immediate\write\psbj@inaux{\lastreadfilename,}%
    \fi%
    \ps@ldinput\lastreadfilename\relax%
    \let\readfilenamehook=\relax%
  }\readfilename%
}%
\expandafter\ifx\csname @@input\endcsname\relax    
  \immediate\let\ps@ldinput=\input\def\input{\psnewinput}%
\else
  \immediate\let\ps@ldinput=\@@input
  \def\@@input{\psnewinput}%
\fi%
\def\nowarnopenout{%
 \def\warnopenout##1##2{%
   \readfilename##2\relax
   \message{\lastreadfilename}%
   \immediate\openout##1=\lastreadfilename\relax}}%
\def\warnopenout#1#2{%
 \readfilename#2\relax
 \def\t@mp{TrashMe,psbjoin.aux,psbjoint.tex,}\uncatcode\t@mp
 \if\matchexpin{\t@mp}{\lastreadfilename,}%
 \else
   \immediate\openin\pst@mpin=\lastreadfilename\relax
   \ifeof\pst@mpin
     \else
     \errhelp{If the content of this file is so precious to you, abort (ie
press x or e) and rename it before retrying.}%
     \errmessage{I'm just about to replace your file named \lastreadfilename}%
   \fi
   \immediate\closein\pst@mpin
 \fi
 \message{\lastreadfilename}%
 \immediate\openout#1=\lastreadfilename\relax}%
{\catcode`\%=12\catcode`\*=14
\gdef\splitfile#1{*
 \readfilename#1\relax
 \immediate\openin\j@insplitin=\lastreadfilename\relax
 \ifeof\j@insplitin
   \message{! I couldn't find and split \lastreadfilename!}*
 \else
   \immediate\openout\j@insplitout=TrashMe
   \message{< Splitting \lastreadfilename\space into}*
   \loop
     \ifeof\j@insplitin
       \immediate\closein\j@insplitin\n@teoffalse
     \else
       \n@teoftrue
       \executeinspecs{\global\read\j@insplitin to\spl@tinline\expandafter
         \ch@ckbeginnewfile\spl@tinline
       \ifc@ntrolline
       \else
         \toks0=\expandafter{\spl@tinline}*
         \immediate\write\j@insplitout{\the\toks0}*
       \fi
     \fi
   \ifn@teof\repeat
   \immediate\closeout\j@insplitout
 \fi\message{>}*
}*
\gdef\ch@ckbeginnewfile#1
 \def\t@mp{#1}*
 \ifx\@mpty\t@mp
   \def\t@mp{#3}*
   \ifx\@mpty\t@mp
     \global\c@ntrollinefalse
   \else
     \immediate\closeout\j@insplitout
     \warnopenout\j@insplitout{#2}*
     \global\c@ntrollinetrue
   \fi
 \else
   \global\c@ntrollinefalse
 \fi}*
\gdef\joinfiles#1\into#2{*
 \message{< Joining following files into}*
 \warnopenout\j@insplitout{#2}*
 \message{:}*
 {*
 \edef\w@##1{\immediate\write\j@insplitout{##1}}*
\w@{
\w@{
\w@{
\w@{
\w@{
\w@{
\w@{
\w@{
\w@{
\w@{
\w@{\string\input\space psbox.tex}*
\w@{\string\splitfile{\string\jobname}}*
\w@{\string\let\string\autojoin=\string\relax}*
}*
 \expandafter\tre@tfilelist#1, \endtre@t
 \immediate\closeout\j@insplitout
 \message{>}*
}*
\gdef\tre@tfilelist#1, #2\endtre@t{*
 \readfilename#1\relax
 \ifx\@mpty\lastreadfilename
 \else
   \immediate\openin\j@insplitin=\lastreadfilename\relax
   \ifeof\j@insplitin
     \errmessage{I couldn't find file \lastreadfilename}*
   \else
     \message{\lastreadfilename}*
     \immediate\write\j@insplitout{
     \executeinspecs{\global\read\j@insplitin to\oldj@ininline}*
     \loop
       \ifeof\j@insplitin\immediate\closein\j@insplitin\n@teoffalse
       \else\n@teoftrue
         \executeinspecs{\global\read\j@insplitin to\j@ininline}*
         \toks0=\expandafter{\oldj@ininline}*
         \let\oldj@ininline=\j@ininline
         \immediate\write\j@insplitout{\the\toks0}*
       \fi
     \ifn@teof
     \repeat
   \immediate\closein\j@insplitin
   \fi
   \tre@tfilelist#2, \endtre@t
 \fi}*
}%
\def\autojoin{%
 \immediate\write\psbj@inaux{\string\into{psbjoint.tex}}%
 \immediate\closeout\psbj@inaux
 \expandafter\joinfiles\GlobalInputList\into{psbjoint.tex}%
}%
%
%
%
\def\centinsert#1{\midinsert\line{\hss#1\hss}\endinsert}%
\def\psannotate#1#2{\vbox{%
  \def\ps@nnotation{#2\global\let\ps@nnotation=\relax}#1}}%
\def\pscaption#1#2{\vbox{%
   \setbox\drawingBox=#1
   \copy\drawingBox
   \vskip\baselineskip
   \vbox{\hsize=\wd\drawingBox\setbox0=\hbox{#2}%
     \ifdim\wd0>\hsize
       \noindent\unhbox0\tolerance=5000
    \else\centerline{\box0}%
    \fi
}}}%
%
\def\at(#1;#2)#3{\setbox0=\hbox{#3}\ht0=0pt\dp0=0pt
  \rlap{\kern#1\vbox to0pt{\kern-#2\box0\vss}}}%
%
\newdimen\gridht \newdimen\gridwd
\def\gridfill(#1;#2){%
  \setbox0=\hbox to 1\pscm
  {\vrule height1\pscm width.4pt\leaders\hrule\hfill}%
  \gridht=#1
  \divide\gridht by \ht0
  \multiply\gridht by \ht0
  \gridwd=#2
  \divide\gridwd by \wd0
  \multiply\gridwd by \wd0
  \advance \gridwd by \wd0
  \vbox to \gridht{\leaders\hbox to\gridwd{\leaders\box0\hfill}\vfill}}%
%
\def\fillinggrid{\at(0cm;0cm){\vbox{%
  \gridfill(\drawinght;\drawingwd)}}}%
%
%
\def\textleftof#1:{%
  \setbox1=#1
  \setbox0=\vbox\bgroup
    \advance\hsize by -\wd1 \advance\hsize by -2em}%
\def\textrightof#1:{%
  \setbox0=#1
  \setbox1=\vbox\bgroup
    \advance\hsize by -\wd0 \advance\hsize by -2em}%
\def\endtext{%
  \egroup
  \hbox to \hsize{\valign{\vfil##\vfil\cr%
\box0\cr%
\noalign{\hss}\box1\cr}}}%
%
\def\frameit#1#2#3{\hbox{\vrule width#1\vbox{%
  \hrule height#1\vskip#2\hbox{\hskip#2\vbox{#3}\hskip#2}%
        \vskip#2\hrule height#1}\vrule width#1}}%
\def\boxit#1{\frameit{0.4pt}{0pt}{#1}}%
\catcode`\@=12 
%
 \psfordvips   

\begin{figure}
$$ \psboxscaled{800}{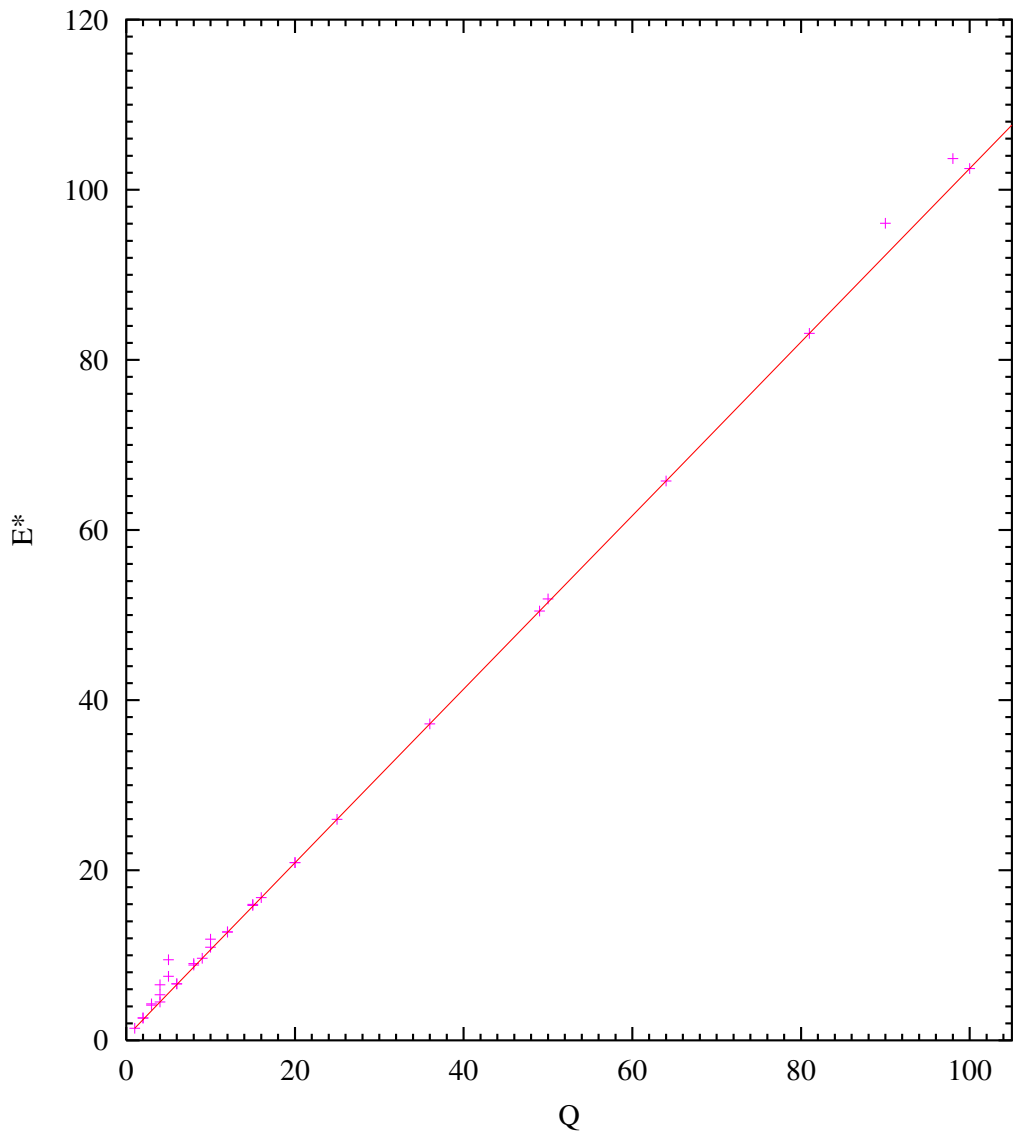} $$
\caption{The rescaled energies $E^* = (32 \pi^2)^{-1} E $ (vertical axis)
are plotted versus the Hopf index $Q=mn$. The straight line connects 
energies for $m=n$. The energies for $m\not= n$ lie slightly above this
line.
}
\end{figure}

\end{document}